\documentclass[aps,prd]{revtex4-1}
\usepackage{amsmath}
\usepackage{amssymb}
\usepackage{amsfonts}
\usepackage{ulem}
\usepackage{graphicx}
\usepackage{times,float}
\usepackage[usenames,dvipsnames,svgnames]{xcolor}
\usepackage{hyperref}
\hypersetup{colorlinks=true, linkcolor=NavyBlue, citecolor=PineGreen,urlcolor=cyan}
\usepackage{multirow}
\usepackage{soul}

\usepackage{ulem}
\usepackage{color}

\newcommand{\beq}{\begin{equation}}
\newcommand{\eeq}{\end{equation}}
\newcommand{\ba}{\begin{eqnarray}}
\newcommand{\ea}{\end{eqnarray}}
\newcommand{\mbf}{\mathbf}
\newcommand{\gv}[1]{\ensuremath{\mbox{\boldmath$ #1 $}}}
\newcommand{\frk}{\mathfrak}
\usepackage{graphicx}
\usepackage[english]{babel}
\usepackage{amsmath,amssymb}
\usepackage{url}
\usepackage{empheq}
\usepackage{epsfig}
\usepackage{subcaption}

\begin{document}

\title{Electromagnetic Radiation in chiral matter: the Cherenkov case }
\author{Eduardo Barredo-Alamilla$^a$}
\email{eduardobarredo@correo.nucleares.unam.mx}
\affiliation{$^a$Instituto de Ciencias Nucleares, Universidad Nacional Aut\'onoma de M\'exico, \\
04510 M\'exico, {Ciudad de M\'exico}, {M\'exico}\looseness=-1}
\author{Luis F. Urrutia$^a$}
\email{urrutia@nucleares.unam.mx}
\affiliation{$^a$Instituto de Ciencias Nucleares, Universidad Nacional Aut\'onoma de M\'exico, \\
04510 M\'exico, {Ciudad de M\'exico}, {M\'exico}\looseness=-1}
\author{Manoel M. Ferreira, Jr.$^b$}
\email{manojr.ufma@gmail.com, manoel.messias@ufma.br}
\affiliation{$^b$Departamento de F\'{i}sica, Universidade Federal do Maranh\~{a}o, Campus
Universit\'{a}rio do Bacanga, S\~{a}o Lu\'is (MA), 65080-805, Brazil}


\begin{abstract}
Starting from  the modified Maxwell equations in Carroll-Field-{Jackiw} electrodynamics we study the electromagnetic radiation in a  chiral medium  characterized by an axion coupling $\theta(x)=b_\mu x^\mu$, with $b_\mu= (0,\mbf{b})$,  which gives rise to the magnetoelectric effect.
Employing the stationary phase approximation we construct the Green's matrix in the radiation zone which allows the calculation of the corresponding electromagnetic potentials and fields for arbitrary sources. We obtain  a general expression for  the angular distribution of the radiated energy per unit frequency. As an application we consider a charge moving
at constant velocity  parallel to $\mbf{b}$ in the medium
and discuss the resulting Cherenkov radiation. We recover the vacuum Cherenkov radiation. For the case of a material with refraction index $n > 1$ we find that  zero, one or two Cherenkov cones can appear. The spectral distribution of the radiation together with the comparison of the radiation output of each cone are presented, as well as some angular plots showing the appearance of the cones.
\end{abstract}

\maketitle

\section{Introduction}

\label{INTRO}

Since its experimental discovery in $1934$ \cite{cherenkov,Vavilov}, Cherenkov
radiation (CHR) has played a fundamental role in the study of the high-energy
particle physics, high-power microwave sources and nuclear and cosmic-ray
physics \cite{Jelley,Jelley1}, both theoretically and phenomenologically. CHR
occurs when charged particles propagate through a dielectric medium with
velocity $v$ higher than $c/\sqrt{\epsilon}$, where $c$ is the speed of
light in vacuum and $\epsilon$ is the permittivity of the medium, which
determines the refractive index $n=\sqrt{\epsilon}$. The first theoretical
description of such radiation in the framework of Maxwell's theory,
developed by Frank and Tamm in Ref. \cite{FT}, reveals its unique
 directional properties. In particular, CHR is produced in a
forward cone defined by the positive angle $\theta= \arccos\left[%
c/\left(vn\right)\right]$ with respect to the direction of the incident
charge. Since the emergence of accelerators in nuclear and high-energy
physics CHR has been widely used to design an impressive variety of
detectors, such as e.g. the ring-imaging \v{C}erenkov detectors \cite%
{Ypsilantis}, which can identify charged particles and also provide a
straightforward effective tool to test its physical properties, like
velocity, energy, direction of motion and charge \cite{PADG}. As remarkable
cases, one mentions that the anti-proton \cite{Chamberlain} and the J-particle \cite{Aubert}
were discovered using CHR detectors.

In this paper we deal with CHR in  chiral media, which belongs to a  class of materials characterized by having a macroscopic electromagnetic response described by non-dynamical  axion electrodynamics \cite{PECCEI,SIKIVIE}. This is an extension of conventional electrodynamics resulting from the addition of the term   $\mathcal{L}_\theta =
(\alpha/4\pi^2)\, {\theta}(x) \, (\mathbf{E}\cdot \mathbf{B})$ to the Lagrangian density \cite{Wilczek,Sekine,Tobar,Paixao}, where $\alpha$ denotes the fine structure constant. Here  $\theta(x)$ is a field known as the axion in particle
physics which we take as a given parameter of the media in the same footing as the permittivity $\epsilon(x)$ and the permeability $\mu(x)$. 
Different materials can be characterized  according to the choice of  $\theta(x)$. For example: (i) magnetoelectric media, that correspond to a non-quantized piecewise constant $\theta(x)$,  were discussed in Refs.\cite{ODELL,Kong,Sihvola,Bianiso,Aladadi,Mahmood,Lorenci,Pedro3,
Urrutia}, (ii) topological insulators \cite{TI1, Urrutia2,Maciejko}, described by a quantized piecewise constant $\theta(x)$, have been also examined in several applied aspects \cite{Chang,Zou, Lakhtakia, Winder,Li,Li1,Ohnoutek,Liang,Tse},  (iii) chiral media, including Weyl semimetals, for instance, which display $\theta(x)=b_\mu x^\mu$ \cite{ARM,Landsteiner,Ruiz,KDeng,Throckmorton,Barnes,Hosur,Day-Nandy}, {and (iv) metamaterials that can realize a synthetic axion response in non-reciprocal artificially designed structures \cite{Shaposhnikov,Prudencio}}.
 The main property  characterizing  these materials is the magnetoelectric effect
(MEE) arising from the additional contribution $\mathcal{L}_\theta$ \cite{QI_SCIENCE}. This coupling produces effective field-dependent charges and
current densities which allow the generation of an electric (magnetic)
polarization due to the presence of a magnetic (electric) field, even in the
static case. Such phenomena were also investigated in the context of the chiral magnetic effect (CME) \cite{Kharzeev1,Kharzeev1B,Fukushima,LiKharzeev,Vilenkin,Inghirami,Chang2,Wurff}, which brings about the magnetic current density $\mathbf{J}_{B}=\sigma_{B}\mathbf{B}$, with $\sigma_{B}$ playing the role of the magnetic conductivity. Classical repercussions arising from  the presence of this current in a dielectric medium were also examined considering symmetric and antisymmetric tensor conductivities \cite{Pedro1}, with the antisymmetric tensor conductivity  further investigated in Ref. \cite{Kaushik1}.

A recently discovered new phenomenon in naturally existing magnetoelectric materials with piecewise constant $\theta(x)$ is the emission of CHR in the backward direction with respect to the incident particle \cite{PRDOJF}.
This remarkable theoretical
prediction is analogue to what occurs in left-handed media (LHM), non-natural materials having simultaneously a negative permittivity and permeability, first examined by Veselago \cite{Veselago}. As these
materials, also called metamaterials, are not readily available in nature,
they have been artificially constructed and tested in the laboratory \cite{Smith-Padilla,Shelby,Parazzoli,Houck,Kadic,Engheta,Pendry,Pendry2}.  Their production has fostered the investigations in backward Cherenkov radiation \cite{LU, Luo, RRCR, Sheng, Chen, Duan,Tao}.

After integration by parts,  the  contribution ${\cal L}_\theta$ of axion-ED with $\theta(x)=b_\mu x^\mu$ reduces to the {$k_{AF}^\mu \sim b^\mu$} term of the CTP-odd contribution of the photon sector in the Standard Model Extension (SME) \cite{KOSTELECKY}, which describes Lorentz invariance violation (LIV) in high energy physics. In this form the theory is also known as the Carroll-Field-Jackiw (CFJ) electrodynamics \cite{CFJ}.  Here, the  vector $b^\mu$  is introduced either as   
an explicit Lorentz violating parameter or as the result  of the spontaneous Lorentz symmetry breaking of a more fundamental theory. 
The dynamics induced by the addition of  ${\cal L}_\theta$ on the usual  Maxwell Lagrangian density  can be also  understood as defining a particular case of  electrodynamics in a medium characterized by  the generalized constitutive relations, $\mbf{D}= \mbf{E}-(\alpha \theta(x)/\pi) {\mbf B}$ and 
$\mbf{H}= \mbf{B}+(\alpha \theta(x)/\pi) {\mbf E}$,  in the case when $\epsilon=1=\mu$.

It is interesting to remark that the CFJ electrodynamics provides an effective theory for condensed-matter systems, accounting for the anomalous Hall effect, the CME, and the electromagnetic response of Weyl semimetals \cite{Yan:2016euz}, which also constitute a typical representative of chiral matter.  In the  latter case one finds that the parameters  $b_0$ and ${ \mathbf b}$ denote the shift in energy and momentum, respectively, of the two Weyl nodes  characterizing the material in the Brillouin zone \cite{BURKOV,FRANZ}. Our $b_\mu$  differs by a factor of two from that in Ref. \cite{FRANZ}. Contrary to the case of LIV in high energy physics, all LIV  parameters in condensed matter physics  arise from the microscopic theory describing the material, thus been well defined and  not necessarily suppressed from an experimental perspective.
  
Vacuum Cherenkov radiation (VCHR) with $n=1$ was discovered in CPT-odd Lorentz-violating theories \cite{POTTING1,POTTING2} and further extended to the CPT-even sector of the SME \cite{ALTSCHUL1, ALTSCHUL2}. Reference \cite{ALTSCHUL1} puts forward the idea that, besides its description as a  conventional radiation process in electrodynamics, 
CHR can be studied from the matter point of view as the decay, $e^{-} \rightarrow e^{-}+ \gamma$, for an arbitrary charge denoted here by $e^-$. The latter approach
was used  to study VCHR for all the Lorentz-violating couplings of fermions that are described by the minimal SME, in the cases with and without spin flip of the fermion \cite{MARCO}. A complete  list of references summarizing  previous works on this topic is also found in this paper. The matter approach  provides a natural way to find subsequent  quantum corrections to the process \cite{SCHWINGER} and it was  used to discuss Cherenkov radiation in the standard vacuum under the influence  of strong electromagnetic fields \cite{MACLEOD}, like those produced by strong laser pulses or in the magnetic field around a pulsar.
 
Further, a fast charged particle passing from a chiral matter to the vacuum emits transition
radiation. Using the matter approach, the photon radiation, $e^- \rightarrow e^-+ \gamma$, and the pair creation, $\gamma \rightarrow e^+ + e^-$, were studied at the boundary between chiral matter and the vacuum \cite{TUCHIN1}. Also the  ultra-relativistic limit for the case of one infinite domain of chiral matter, together with  the case of two semi-infinite domains separated by a domain wall  was considered in Ref. \cite{TUCHIN2}. The main features of the radiation were shown to depend on the parameters of the chiral anomaly in these materials.  Also the high energy limit of the frequency spectrum and the angular distribution for Cherenkov radiation in chiral matter was obtained \cite{TUCHIN3}. Anomalous scattering of fermions in matter induced by the chiral anomaly was also investigated, concluding that the scattering angles are proportional to the chiral conductivity \cite{TUCHIN4}. Recent studies on collisional energy loss and bremsstrahlung in chiral medium have also been reported with great interest \cite{JHansen}.

Our discussion of the CHR in a chiral media  parallels the  analysis of  radiation by charged particles  described  in regular references of electrodynamics \cite{Schwinger,JACKSON, Panofsky}. We obtain a general formula for the radiation fields produced by arbitrary sources, which can be subsequently used to determine all the relevant observable quantities  in the radiation processes. As an application of these general findings we next concentrate in  CHR, 
i.e. we consider a charge $q$ moving at constant velocity in the medium, ${|\mbf v|> c/n}$,  and  neglect recoil effects. We do not restrict ourselves to the ultra high energy limit, thus allowing to address the whole range of (charge) velocities,  $  |{\mbf v}|/n < |{\mbf v}| <c $. Also, we consider a material with $b_0=0$ and choose the charge velocity parallel to ${\mbf b}$, in such a way to assure axial symmetry in our model.

The paper is organized as follows. In section \ref{CHED} we summarize the main aspects of CFJ electrodynamics which we use in the following. Section \ref{GF} is devoted to the construction
of the Green's function (GF) of the system in momentum and coordinate space.  To this end we  further extend to the time-dependent case the methods  already developed in Refs. \cite{Ruiz,UrrutiaMartinCambiaso3,UrrutiaMartinCambiaso,
UrrutiaMartinCambiaso4} for
the static case. In section \ref{GFRAD} we  adopt the stationary phase approximation to calculate the GF in  the  radiation zone. To make the subsequent calculations still analytic we have to introduce  a further approximation in the solution of the stationary phase equation, which determines the range of validity in our calculations that  must be verified  after obtaining the final results. The resulting electromagnetic fields in the radiation zone are presented  in section \ref{EMFRAD} for an arbitrary current $J^\mu$. 
Our calculations up to  section \ref{EMFRAD} together with the Apendixes \ref{APPA}, \ref{APPB} and \ref{APPC} are for  $n=1$.
An arbitrary index of refraction $n$ is  introduced in  sections \ref{CHRAD}  and  \ref{TOTRADEN} following the transformations indicated in the Appendix \ref{APPD}.
In section \ref{CHRAD} we consider  the particular case of CHR and  we calculate the electromagnetic fields,
the angular distribution of the radiated energy per unit frequency, the Cherenkov angles and their behavior as a function of the parameter $|{\mbf b}|$.  The total radiated power per unit frequency is discussed in section \ref{TOTRADEN}, where the ratio between the radiation output of  the two possible cones in chiral matter is calculated, as well as the ratio of the production of  each of them with respect to the conventional case. Section \ref{SUM} contains the summary and conclusions. Besides the Appendix \ref{APPD} already mentioned, the details of the calculation of the GF for $n=1$ in the radiation zone are summarized in the Appendix \ref{APPA}. A general expression for the spectral distribution of the radiation for arbitrary sources and $n=1$ is obtained in the Appendix \ref{APPB}, which also summarizes the symmetry properties of some auxiliary functions  introduced in the text. The main steps in the calculation for the spectral distribution in the particular case of CHR is  presented in the Appendix \ref{APPC} for $n=1$. Let us emphasize again that any result obtained for $n=1$ can be generalized to arbitrary $n$ following the prescriptions indicated in the Appendix \ref{APPD}.

\section{CFJ Electrodynamics}

\label{CHED}

In terms of the
electromagnetic potential, $A_\mu=(\Phi, -\mathbf{A})$, the action is 
\begin{equation}
S[A_\mu(x)] =\int d^4x \left[-\frac{1}{16 \pi} F_{\mu\nu}F^{\mu\nu} -\frac{1}{c}J^\mu A_\mu- 
\frac{\alpha}{16\pi^2}{\theta}(x)F_{\mu\nu}\tilde{F}^{\mu\nu}\right],  \label{ACTION}
\end{equation}
where $F_{\mu\nu}=\partial_\mu A_\nu- \partial_\nu A_\mu$ and $\tilde{F}^{\mu\nu}=\frac{1}{2}\epsilon^{\mu\nu\rho%
	\sigma}F_{\rho\sigma}$ are the
electromagnetic field strength and its dual tensor, respectively, $J^\mu=(c \rho, {\mathbf J}) $ is a conserved external current while $\theta(x)=b_\mu x^\mu$ is the axion coupling. As usual, the electromagnetic fields are 
\begin{equation}
\mbf{B}= \gv{\nabla}\times \mbf{A}, \qquad \mbf{E}=-\gv{\nabla} A^0- \frac{1}{c} \frac{\partial \mbf{A}}{\partial t}.
\end{equation}
Our metric is $%
\eta^{\mu\nu}=\mathrm{diag}(1,-1,-1,-1)$, we set $\epsilon^{0123}=+1$ and employ the unrationalized Gaussian units, following the conventions of Ref. \cite{JACKSON}.

 For a general axion field $\theta(x)$ the action (\ref{ACTION}) violates translation invariance and Lorentz symmetry \cite%
{KOSTELECKY}, but it is manifestly gauge invariant. Nevertheless, in the particular  case considered
where $b_\mu$ is a constant vector, we still have translational invariance. In fact, under the  translation, $x^\mu \rightarrow x^\mu+ a^\mu$ the additional term  proportional to $(b_\mu a^\mu) F_{\mu\nu}{\tilde F}^{\mu\nu}$  appears in the action (\ref{ACTION}). Recalling that $ F_{\mu\nu}{\tilde F}^{\mu\nu}$  is a total derivative, the invariance of the action under translations is thus recovered. An equivalent way to verify the invariance under translations  is 
integrating by parts the last term in Eq.(\ref{ACTION}), which produces the Carroll-Field-Jackiw electrodynamics \cite{CFJ}, 
\begin{equation}
	S_{MCFJ} =\int d^4x \left[-\frac{1}{16 \pi} F_{\mu\nu}F^{\mu\nu} -\frac{1}{c}J^\mu A_\mu+ 
	\frac{\alpha}{16\pi^2}b_\mu A_\nu \tilde{F}^{\mu\nu}\right],  \label{ACTIONCFJ}
\end{equation}
where the translation invariance is manifest and the gauge invariance is granted only up to a total derivative, however. As expected, the resulting equations of motion are  gauge invariant, being
\begin{eqnarray}
&&\partial_\mu F^{\mu\nu}= \frac{4\pi}{c} J^\nu - {\tilde b}_\mu \tilde{F}^{\mu\nu}, \qquad {\tilde b}_\mu =
\frac{\alpha}{\pi} b_\mu,
\label{MODM}
\end{eqnarray}
which read 
\begin{eqnarray}
[\eta^{\mu \nu} \partial^2- \partial^\mu\partial^\nu -
\epsilon^{\mu\nu\rho\sigma} {\tilde b}_\rho \partial_\sigma]A_\nu = J^\mu,  \label{MODA}
\end{eqnarray}
in terms of the potential.
The inhomogeneous Maxwell's equations are
\begin{eqnarray}
&&\gv{\nabla} \cdot \mathbf{E} = 4\pi \rho-\gv{\tilde{b}}\cdot \mathbf{B}, \qquad \gv{\nabla}
\times \mathbf{B} -\frac{1}{c}\frac{\partial \mathbf{E} }{\partial t} = \frac{4\pi}{c}\mathbf{J}+{\tilde{b}}%
_0 \mathbf{B} + \gv{\tilde{b}}\times\mathbf{E}.  \label{MAX1}
\end{eqnarray}
We still have the homogeneous Maxwell equations arising from the Bianchi
identity, $\partial_\mu \tilde{F}^{\mu\nu}=0$,
\begin{equation}
\gv{\nabla} \cdot \mathbf{B} = 0, \qquad \gv{\nabla} \times \mathbf{E} +\frac{1}{c}\frac{\partial \mathbf{B}
}{\partial t}  = 0.  \label{MAX2}
\end{equation}

In Eqs. (\ref{MAX1}) we have $\gv{\tilde{b}}=\mathbf{\nabla }\theta$ and ${\tilde{b}}_0=\partial _{t}\theta$, where $\theta$ is the axion field. The terms involving derivatives of  $\theta$  play relevant roles in condensed matter systems \cite{ZQ,Pedroo}. Indeed, while $\mathbf{\nabla }\theta \cdot \mathbf{B}$ represents an anomalous charge density, $\mathbf{\nabla }\theta \times \mathbf{B}$ appears in the anomalous Hall effect (AHE) and  $(\partial _{t}\theta )\mathbf{B}$ stands for the chiral magnetic current \cite{Kharzeev1,Kharzeev1B,Fukushima,LiKharzeev,Vilenkin,Inghirami,Chang2,Wurff,Pedro1,Kaushik1}. In the case when the
axion field does not depend on the space coordinates, $\mathbf{\nabla }\theta
={\bf{0}},$   the Maxwell equations (\ref{MAX1}) read
\begin{equation}
	\mathbf{\nabla }\cdot \mathbf{E}=4 \pi \rho, \qquad
	\mathbf{\nabla }\times 
	\mathbf{B}-\frac{1}{c}\frac{\partial \mathbf{E}}{\partial t}=\frac{4 \pi}{c}\mathbf{J}+(\partial _{t}\theta )\mathbf{B},
	\label{Maxwellaxion1}
\end{equation}
with $\partial _{t}\theta$ representing the magnetic conductivity, $\sigma_{B}$, in the chiral current, $\mathbf{J}_{B}=\sigma_{B}\mathbf{B}$. Effects of the CFJ term, together with those arising from its higher derivative dimension five counterpart, on the electromagnetic propagation in continuous matter were  analyzed from a classical perspective in Ref. \cite{Pedroo}.

Since our main interest is in radiation processes, we look for  the energy density ${{\bar u}}$ and the energy flux $\mathbf{\ {\bar S}}\,$ satisfying the  conservation equation $\partial_t {\bar u}+ \gv{\nabla}\cdot {\mathbf {\bar S}}=0$ outside the sources.  From the standard manipulations of Maxwell's equations (\ref{MAX1}) and (\ref{MAX2}) we obtain
\begin{equation}
-{\mathbf J}\cdot {\mathbf E}= \frac{\partial u}{\partial t}+ \ensuremath{\mbox{\boldmath$ \gv{\nabla} $}}%
\cdot \mathbf{\ S}+\frac{c}{4\pi} \tilde{b}_0 (\mathbf{E}\cdot \mathbf{B}),
\label{NCONS}
\end{equation}
with 
\begin{equation}
{u}= \frac{1}{8 \pi} \Big( \mathbf{E}^2 + \mathbf{B}^2 \Big), \qquad \mathbf{S}= \frac{c}{4 \pi}
\mathbf{E}\times \mathbf{B}.
\label{US}
\end{equation}
The identity
\begin{equation}
\mathbf{E}\cdot\mathbf{B}= -\frac{1}{2c} \frac{\partial}{\partial t} (\mbf{A}\cdot\mbf{B})+ \frac{1}{2}\gv{\nabla}\cdot (\mbf{A}\times\mbf{E}-A^0 \mbf{B}),
\end{equation}
allows us to define the  energy density $\bar{u}$, together with the corresponding Poynting vector $\mbf{\bar S}$, as
\begin{equation}
{\bar u}= u- \frac{1}{8\pi}{\tilde b}_0 \mbf{A}\cdot {\mbf B}, \qquad \mbf{\bar S}=\mbf{S}+\frac{1}{8\pi}{\tilde b}_0 (\mbf{A}\times \mbf{E}-A^0 \mbf{B}), \label{MODUS}
\end{equation}
fulfilling the required conservation equation when ${\mathbf J}=0$. Similar results are obtained from the covariant version of the energy-momentum tensor in Refs. \cite{CFJ,POTTING1,POTTING2}, which confirms $\bar u$ and $\bar S$, given in Eq. (\ref{MODUS}), as the energy density and and energy flux that respect the continuity equation.

Let us emphasize that under the gauge transformation, $ \delta_\Lambda A^0 =\frac{1}{c} \frac{\partial \delta \Lambda}{\partial t}, \, \delta_\Lambda \mbf{A} =  - \gv{\nabla} \delta \Lambda$, the terms that depend on the potential $A^\mu$ in Eq. (\ref{MODUS}) are gauge invariant up to a total derivative, changing  as
\begin{equation}
\delta_\Lambda (\mbf{A}\cdot\mbf{B}) = \Lambda \gv{\nabla}\cdot\mbf{B},\qquad 
\delta_\Lambda(\mbf{A}\times \mbf{E}-A^0 \mbf{B}) =\lambda \left(\gv{\nabla}\times \mbf{E}-\frac{1}{c}\frac{\partial \mbf{B}}{\partial t}\right), 
\end{equation}
which yield  null results when one takes into account the homogeneous Maxwell equations.
From Eq. (\ref{MODUS}) we realize that $\bar{u}$  is not positive definite, which prompts us to set  $b_0=0$ in the following to avoid instabilities in the system \cite{ZQ}. Also, we take the $z$-axis in the direction of the $\mbf{b}$ vector

\section{The Green's function}
\label{GF}

\subsection{Green's function in momentum space}

\label{GFMOM}

Our next step is to construct the Green function (GF) $G_{\mu\nu}(x- x^{\prime})$
of CFJ electrodynamics in the time dependent case. It is defined by 
\begin{equation}
[\eta^{\mu \nu} \partial^2- \partial^\mu\partial^\nu -{\tilde b}_\rho
\epsilon^{\mu\nu\rho\sigma} \partial_\sigma] G_{\nu \beta}(x-x^{\prime})=
 \delta^\nu{}_\beta \delta^4(x-x^{\prime}).
\label{EQGF}
\end{equation}
Going to momentum space, we write 
\begin{equation}
G_{\nu \beta}(x-x^{\prime})=\int \frac{d^4 k}{(2 \pi)^4} 
e^{-ik_\mu{(x-x^{\prime})^\mu}}G_{\nu \beta} (\omega, 
\mathbf{k}),\quad x^\mu=(ct, \mbf{x}),  \quad k^\mu= (\omega/c, \mathbf{k}), \quad \partial_\mu=-i k_\mu, \label{DEFMOM}
\end{equation}
obtaining 
\begin{equation}
[-k^2 \eta^{\mu \nu} + k^\mu k^\nu +i\epsilon^{\mu\nu\rho\sigma}{\tilde b}%
_\rho k_\sigma]G_{\nu\beta}(k) = \delta^\mu{}_\beta, 
\label{MOMREP}
\end{equation}
which in the Lorentz gauge, $\partial_\mu A^\mu=0$, reduces to 
\begin{equation}
[-k^2 \eta^{\mu \nu} +i\epsilon^{\mu\nu\rho\sigma}{\tilde b}_\rho
k_\sigma]G_{\nu\beta}(k) = \delta^\mu{}_\beta,
\label{LG}
\end{equation}
with $k^2= k_0^2-\mathbf{k}^2$. A long but straightforward calculation yields 
\begin{equation}
G_{\nu \lambda}(k) = -\frac{k^2 \eta_{\nu \lambda} + {\tilde b}_\nu {\tilde b%
}_\lambda + i \epsilon_{\nu \lambda \alpha \beta} {\tilde b}%
^{\alpha}k^{\beta} }{k^4 +{\tilde b}^2 k^2 - ({\tilde b} \cdot k )^2}+ \frac{%
({\tilde b}\cdot k)({\tilde b}_\lambda k_\nu+{\tilde b}_\nu k_\lambda)- {%
\tilde b}^2k_\nu k_\lambda}{k^2(k^4 -{\tilde b}^2 k^2 + ({\tilde b} \cdot k
)^2)} .  \label{GFLG}
\end{equation}
At this stage we can verify that $k^\nu G_{\nu\lambda}(k)=0$. The above
expression can be further simplified recalling the relation $A_\nu (k) =
G_{\nu \lambda}(k) J^{\lambda}(k) $. Since $G_{\nu \lambda}(k)$ couples to a
conserved current, we can dispose of all factors proportional to $k_\lambda J^{\lambda}$.
Furthermore, note that any contribution proportional to $k_{\nu}$ corresponds to a gauge
transformation in the resulting $A_\nu$. Then, a simpler representation of
the function (\ref{GFLG}), without loss of generality, is 
\begin{equation}
G_{\nu \lambda} = -\frac{k^2 \eta_{\nu \lambda} + {\tilde b}_\nu {\tilde b}%
_\lambda + i \epsilon_{\nu \lambda \alpha \beta} {\tilde b}%
^{\alpha}k^{\beta} }{k^4 +{\tilde b}^2 k^2 - ({\tilde b} \cdot k )^2},
\label{GFSIMP}
\end{equation}
coinciding with the result of {Refs. \cite{POTTING1,POTTING2, ZQ} }, but no longer  written in the {Lorenz} gauge.

\subsection{Green's function in coordinate space}

\label{GFCOORD}

In order to deal with radiation, we need to express the GF in the coordinate
space. To this end, we keep the dependence on the frequency and perform the
Fourier transform only in coordinate space. We set ${\tilde{b}}_{\mu}=(0,0,0,b),$ with $b=|\gv{\tilde{b}}|= \alpha |\mbf{b}|/\pi$. Starting from Eq. (\ref{GFSIMP}), we write
\begin{eqnarray}
G^{\mu \nu }(\mathbf{x},\mathbf{x^{\prime }};\omega ) &=&\int \frac{d^{3}k}{%
(2\pi )^{3}}G^{\mu \nu }(\omega ,\mathbf{k})e^{i\mathbf{k}\cdot (\mathbf{x}-%
\mathbf{x^{\prime }})},  \notag \\
G^{\mu \nu }(\mathbf{x},\mathbf{x^{\prime }};\omega ) &=&-\int \frac{d^{3}k}{%
(2\pi )^{3}}\frac{k^{2}g^{\mu \nu }+ib\epsilon ^{\mu \nu 3\sigma }k_{\sigma
}+{{\tilde{b}}}^{\mu }{{\tilde{b}}}^{\nu }}{(k^{2})^{2}+{{\tilde{b}}}%
^{2}k^{2}-({{\tilde{b}}}\cdot k)^{2}}e^{i\mathbf{k}\cdot (\mathbf{x}-\mathbf{%
x^{\prime }})},  \notag \\
G^{\mu \nu }(\mathbf{x},\mathbf{x^{\prime }};\omega ) &=&-\left[ (k_0
^{2}+\gv{\nabla} ^{2})g^{\mu \nu }+ik_0 b\epsilon ^{\mu \nu 30}-b\epsilon
^{\mu \nu 3i}\partial _{i}+{\tilde{b}}^{\mu }{\tilde{b}}^{\nu }\right] f(%
\mathbf{x},\mathbf{x^{\prime }};\omega ),  \label{FGCOMPL1}
\end{eqnarray}%
in terms of  $\partial _{j}=-ik_{j}$, and introduce the
function,
\begin{equation}
f(\mathbf{x},\mathbf{x^{\prime }};\omega )=\int \frac{d^{3}k}{(2\pi )^{3}}%
\frac{e^{i\mathbf{k}\cdot (\mathbf{x}-\mathbf{x^{\prime }})}}{(k^{2})^{2}+{%
\tilde{b}}^{2}k^{2}-({\tilde{b}}\cdot k)^{2}},  \label{F0}
\end{equation}%
from which we calculate the Green's function according to Eq. (\ref%
{FGCOMPL1}). Under our conventions, the denominator in Eq. (\ref%
{F0}) is 
\begin{equation}
D\equiv (k^{2})^{2}+\tilde{b}^{2}k^{2}-(\tilde{b}\cdot k)^{2}=(k_0^{2}-%
\mathbf{k}^{2})^{2}-b^{2}(k_0^{2}-\mathbf{k}^{2})-b^{2}k_{3}^{2}. \label{DEN}
\end{equation}
The condition $D=0$ yields the dispersion relation 
\beq
k_0^2=k_\perp^2+ \left(\sqrt{k_z^2+ b^2/4} \pm b/2 \right)^2,\label{DR}
\eeq
where
\begin{equation}
{\mbf k}^2= k_\perp^2+ k_z^2, \quad k_\perp=+\sqrt{ k_x^2+k_y^2}.
\end{equation}
Setting $b=0$ in the dispersion relation (\ref{DR}), one recovers the conventional  vacuum result.

The evaluation of Eq. (\ref{F0}) is performed in cylindrical coordinates using the integration over the polar angle to introduce the Bessel function
$J_0$ and, subsequently, calculating the integral over  $k_z$ in the complex plane.
To this end, we rewrite the denominator $D$ as
\beq
D=\left(k_z^2-(k_z^-)^2 \right)\left(k_z^2-(k_z^+)^2 \right),
\label{DEN1}
\eeq
which allows the identification of the poles in $k_z$, 
\begin{equation}
k_z^{\pm}= \sqrt{k_\parallel^2 \pm b k_\parallel},
\label{DEFKPAR1}
\end{equation}
with the redefinition,
\begin{equation}
k_\parallel=\sqrt{k_0^2-k_\perp^2}.
\label{DEFKPAR2}
\end{equation}
The integration over $k_z$ is long but straightforward, yielding the final result 
\begin{equation}
f(\mathbf{x},\mathbf{x^{\prime }};\omega )=\frac{i}{8\pi }\int_{0}^{\infty } 
\frac{k_{\perp} \, dk_{\perp}}{k_\parallel} \,J_{0}\Big(R_{\perp } k_\perp%
\Big)\frac{1}{b}\left[ \frac{e^{i\sqrt{k_{\parallel }^{2}+bk_{\parallel }}Z}%
}{\sqrt{k_{\parallel }^{2}+bk_{\parallel }}}-\frac{e^{i\sqrt{k_{\parallel
}^{2}-bk_{\parallel }}Z}}{\sqrt{k_{\parallel }^{2}-bk_{\parallel }}}\right],
\label{FCOMPL}
\end{equation}
where we stick to our independent
variable $k_\perp > 0$ and define $Z=|z-z'|$. 
Setting $k_0 =0$, we recover the static approximation considered in Ref. \cite{ZQ}.

\section{The Green's function in the radiation zone}

\label{GFRAD}

\begin{figure}[h!]
\centering 
\includegraphics[scale=0.5]{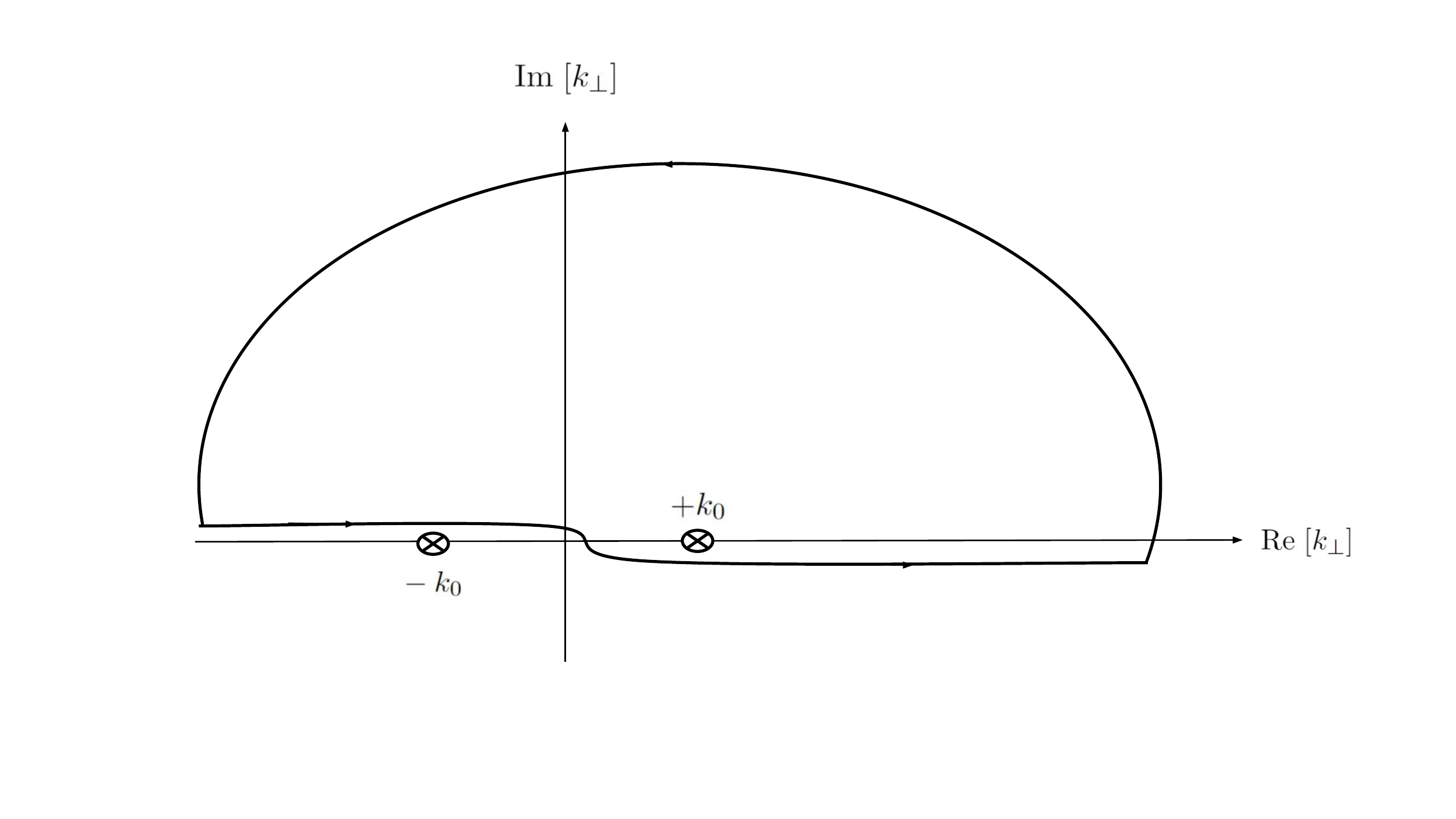} %
\caption{The Sommerfeld integration path. The path is above the logarithmic branch point singularity of $H_0^{(1)}(K_\perp R_\perp)$ at $k_\perp =0$. Furthermore $k_\parallel=\sqrt{k_0^2-k_\perp^2}$ has singularities at $k_\perp=\pm k_0$.}
\label{FIG11}
\end{figure}

The relations,
$J_{0}(x)=(H_{0}^{(1)}(x)+(H_{0}^{(2)}(x))/2$ and $
H_{0}^{(1)}(e^{i\pi }x)=-H_{0}^{(2)}(x)$,
allow to extend the integration limit of $k_\perp$ from $-\infty $ to $
+\infty$, yielding the convenient expression 
\begin{equation}
f(\mathbf{x},\mathbf{x^{\prime }};\omega )=\sum_{\eta=\pm 1} f_{\eta}(\mathbf{x%
},\mathbf{x^{\prime }};\omega ), \quad f_\eta (\mathbf{x},\mathbf{x^{\prime }};\omega )=\frac{i}{16\pi }%
\int_{-\infty }^{\infty }\frac{k_{\perp} \, dk_{\perp}}{k_\parallel}%
\,H_{0}^{(1)}\Big(k_\perp R_{\perp }\Big)\frac{1}{\eta b }\frac{e^{i\sqrt{%
k_{\parallel }^{2}+\eta b\, k_{\parallel }}Z}}{\sqrt{k_{\parallel }^{2}+\eta
b\, k_{\parallel }}},  \label{FETA}
\end{equation}%
with $b>0$ everywhere, and the sum on $\eta=\pm 1$ contemplates the two square roots present in Eq. (\ref{FCOMPL}). The extension to the complex plane of the integral on the variable $k_\perp$ is made by introducing the Sommerfeld path, shown schematically  in Fig. \ref{FIG11}.

In the following we concentrate upon the calculation of $f_{\eta}(\mathbf{x},\mathbf{x^{\prime }%
};\omega )$ in the far-zone regime, $r=|{\mathbf{x}}| \gg |{\mathbf{x^{\prime }}}|=r'$, where we have
highly oscillating functions in the integrand of 
Eq.\ (\ref{FETA}). This property suggests the use of
the stationary phase approximation (SPA) to evaluate the integral. We consider the approximation  $R_{\perp}=|(\mathbf{x}-\mathbf{x^{\prime })}_{\perp }|= R |\sin \theta|$, together with  $Z=|x_{3}-x_{3}^{\prime}|= R |\cos \theta|$, where 
$R=|(\mathbf{x}-\mathbf{x^{\prime })}|$ and $\theta$ is the polar angle of the observation point determined by $\mbf{x}$. To assess the validity of the approximation we have to compare the directions of the vectors $\mbf{x}$ and $\mbf{R}=\mbf{(x-x')}$, with respect to the $z$-axis. In the most unfavorable situation, when $\mbf{x'}$ is orthogonal to $\mbf{x}$, we can show that the angle $\Theta$ that the vector ($\mbf{x}-\mbf{x'}$) makes with the $z$-axis is such that $\cos \Theta= \cos \theta-(r'/r) \sin \theta$, to first order in $r'/r$. That is to say, it is a very good approximation to take $\Theta \approx \theta$ in the regime $r'/r \ll 1$. This, together with the asymptotic behavior
\begin{eqnarray}
&&H_{0}^{(1)}(k_{\perp }R_{\perp })= \sqrt{\frac{2}{\pi k_{\perp
}R_{\perp }}}e^{ik_{\perp }R_{\perp }-i\pi /4}, \label{CONST}
\end{eqnarray}%
yields
\begin{equation}
f_{\eta }(\mathbf{x},\mathbf{x^{\prime }};\omega )=\frac{ie^{-i\pi /4}}{%
16\pi }\sqrt{\frac{2}{\pi }}\int_{-\infty }^{\infty }\frac{k_{\perp
}\,dk_{\perp }}{k_{\parallel }}\,\frac{1}{\sqrt{k_{\perp }r\sin \theta }} 
\frac{1}{\eta b} \frac{e^{iR\left( k_{\perp }|\sin \theta| +\sqrt{k_{\parallel
}^{2}+\eta b\,k_{\parallel }}|\cos \theta| \right) } }{\sqrt{k_{\parallel
}^{2}+\eta b\,k_{\parallel }}}.  \label{OSC_PHAS}
\end{equation}
As usual, we further replace $R$ by $r$, except in the phase of the exponential, where we assume $R= \left( r-\hat{\mathbf{n}}\cdot \mathbf{x^{\prime }}\right)$, with $\mbf{\hat{n}}=\mbf{x}/r$, thus taking into account the phase modifications induced by the source.

\begin{figure}[htb!]
\includegraphics[width=0.45\linewidth]{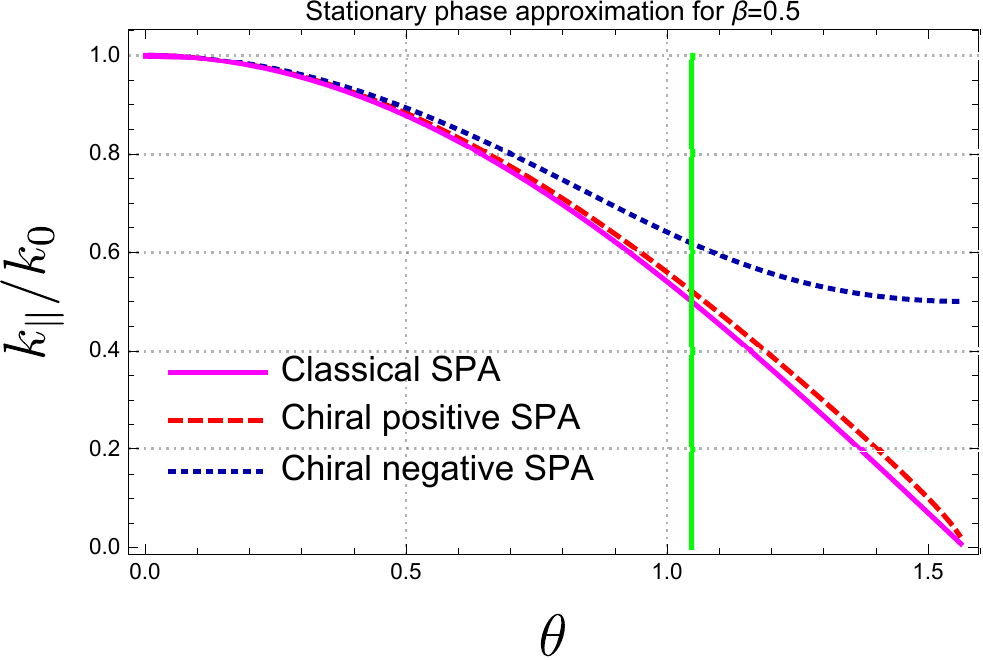}
\hspace{0.5cm}
\includegraphics[width=0.45\linewidth]{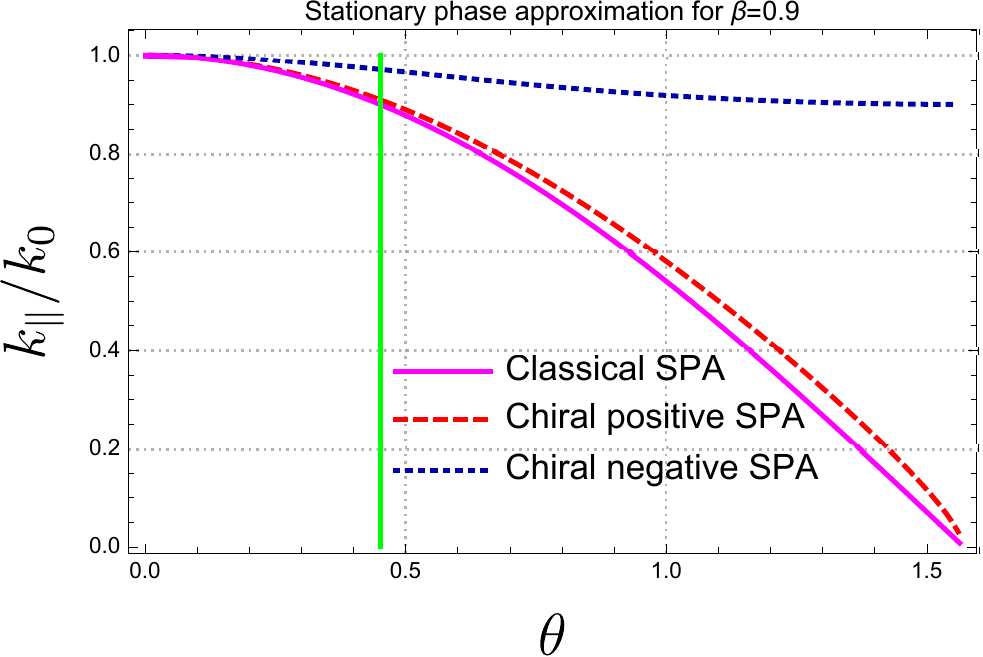}
\caption{The exact numerical solution $\kappa(\theta)$ of Eq. (\ref{SPA}) is indicated  with the dashed (red) line for $\eta=+1$ and with the dotted (blue) line for $\eta=-1$. The solution $\kappa(\theta)$ of the approximation in Eq. (\ref{APPK}), called classical SPA, is plotted as a solid (magenta) line.
 The vertical solid (green) line indicates the angle $\theta=\arccos \alpha$,  to the right of which $\sqrt{k^2_\parallel-b k_\parallel}$ becomes imaginary. Left panel:   $\beta=0.5$. Right panel: $\beta= 0.9$  } 
\label{FIG01}
\end{figure}

We recall the general expression for the SPA
\begin{equation}
I=\int_{-\infty }^{+\infty }\,dt\,e^{iR \,h(t)}\,f(t)=e^{iR\, h(t_{0})}\,f(t_{0})\,%
\sqrt{\frac{2\pi i}{rh^{\prime \prime }(t_{0})}},  \label{SPA0}
\end{equation}%
where $R\gg 1$ and $t_{0}$ is the solution of $h^{\prime }(t)=0$, which makes $h^{\prime
}(t_{0})$ an extremum of $h(t).$ The prime ($^\prime$) denotes the derivatives with
respect to $t$ in the usual fashion. The functions $h(t)$ and $f(t)$ are
identified by comparing with those in  Eq.\ (\ref{OSC_PHAS}) after the far field
approximation is imposed. In Eq.\ (\ref{OSC_PHAS}), we choose to find the stationary phase by looking at the extreme of the functions,
\begin{equation}
h_\eta(k_{\perp })= k_\perp  |\sin \theta|+ \,\sqrt{%
k_{\parallel }^{2}+\eta b \,k_{\parallel }}\, |\cos \theta|>0. \label{PHASE}
\end{equation}%
To this end, we take the derivative of $h_\eta$ with respect to $k_{\parallel}$ for simplicity. Since $d k_\parallel/d k_\perp \neq 0 $ and the resulting relation is set equal to zero, the outcome is independent of the variable chosen to calculate the derivative. After rearranging the result of $d h_\eta(k_\perp)/dk_\parallel$, we find it convenient to present the  condition for the extremum  as 
\begin{equation}
{\frac{\kappa}{\sqrt{1-\kappa ^{2}}}\tan \theta }=\left(1 +\frac{\eta}{2}\frac{\beta}{\kappa} \right)
\left(1 +\eta\frac{\beta}{\kappa} \right)^{-1/2},
\label{SPA}
\end{equation}%
in terms of the dimensionless variables $\kappa $ and $\beta $, defined as
\begin{equation}
\kappa =\frac{k_{\parallel }}{k_0 }>0,\qquad \beta =\frac{b}{k_0 } > 0.  \label{SPA2}
\end{equation}
The Eq. (\ref{SPA}) is a quartic equation for $\kappa $ which {is difficult to solve analytically}. For a given $\beta$, the exact numerical solution $\kappa(\theta)$ of Eq. (\ref{SPA}) is indicated  with the dashed (red) line for $\eta=+1$ and with de dotted (blue) line for $\eta=-1$, in the panels of Fig. \ref{FIG01}.  To proceed further  we resort to an approximation in the solution of Eq.\ (\ref{SPA}) considering the limit $b \ll k_\parallel < k_0$, where {$\beta/\kappa \ll 1$}. Then, Eq. (\ref{SPA}) simplifies as 
\beq
\frac{\kappa ^{2}}{1-\kappa ^{2}}\tan ^{2}\theta =1 + {\frac{1}{4}}
\left(\frac{\beta}{\kappa} \right)^2 +O(\beta^3).
\label{SPAAPP}
\eeq 
To first order in  ${\beta}/{\kappa} $, we obtain  $\kappa= \cos \theta$ as our approximate SPA solution, for each value of $\eta$.  This choice yields
\beq
k_\parallel(\theta, \eta)= k_0 |\cos \theta|, \qquad 
k_\perp(\theta, \eta)= k_0 |\sin  \theta|. \label{APPK}
\eeq
The solution (\ref{APPK}), called the classical SPA, is plotted as the solid (magenta) line in Fig. \ref{FIG01}.  Let us emphasize that this approximation is taken only to have { a simpler} analytic way to proceed with the calculation and that $b,  (\beta)$, is taken non-zero in all the required remaining functions. The approximation is valid whenever $\cos \theta \gg \beta/(2 \sqrt{2})$, which should be verified  at the end of any  determination of a Cherenkov angle. In any case, Figs. \ref{FIG01} provide a qualitative idea of the validity of the approximation (\ref{APPK}), showing that it is very good for the contribution $\eta=+1$ in the whole angular range. Looking back to Eq. (\ref{FETA}) we realize that this is not the case for the contribution $\eta=-1$, which is
 suppressed to the right of the vertical (green) line shown in the figures, where  $\sqrt{k^2_\parallel-b k_\parallel}$ becomes imaginary. This line corresponds to $\theta_0= \arccos \beta$.

Recalling Eqs. (\ref{OSC_PHAS}) and  (\ref{SPA0}), and substituting  the SPA  (\ref{APPK}), we arrive at  
\ba
&&  f(\mbf{x},\mbf{x'};\omega)|_{\rm SPA}\equiv \frk{f}(\mbf{x},\mbf{x'};\omega),  \nonumber \\
&&\frk{f}(\mbf{x},\mbf{x'};\omega)  = \frac{1}{8\pi r}  \frac{1}{k_0 \, b |\cos\theta|} \sum_{\eta=\pm 1}   \frac{e^{ik_0  C_\eta(\theta)
 (r-\mbf{\hat{n}}\cdot \mbf{x'}) }}{ g_\eta(\theta) },
 \label{FFIN}
\ea
with
  
\ba
&&\hspace{-0.5cm} C_\eta(\theta)=\left(\sin^2\theta+ \cos^2\theta \sqrt{1+ \eta \beta |\sec\theta|} \right),  \label{CETA}\\
&& \hspace{-0.5cm}  g_\eta (\theta)=\frac{1}{\left( 1+\eta \beta |\sec \theta| \right)^{1/4}}
\sqrt{ \left(1+ \eta \beta |\sec\theta|\right) \left(1 +\frac{\eta \beta}{2} |\sec\theta|\right) + \frac{\beta^2}{4}{\tan^2\theta}}.
\label{GETA}
\ea
Let us remark the following obvious symmetry properties in terms of the frequency
\beq
C_\eta(-\omega)= C_{-\eta}(\omega), \qquad  g_\eta(-\omega)= g_{-\eta}(\omega),
\label{SYMCG}
\eeq
which will be useful in the following.
A word of caution is required here. If we had solved the stationary phase equation exactly, any change of variables in the subsequent integration would produce  the same result. This is not the case when using an approximation as we have done. Since we have chosen $k_\perp$ as our integration variable, we have to make sure that the second derivative $h''_\eta$ is calculated with respect to $k_\perp$. 
Notice that in the limit  $\alpha=0$ we have $C_\eta(\theta)=1$, and we recover the phase $ik_0 R$ 
describing conventional radiation. Also, $g_\eta(\theta)=1$ in that limit.

We choose to split the GF in Eq. (\ref{FGCOMPL1}) 
as
\begin{eqnarray}
G_{0}^{\mu \nu }(\mathbf{x},\mathbf{x^{\prime }};\omega ) &=&-g^{\mu \nu
}(k_0^{2}+\gv{\nabla} ^{2})f(\mathbf{x},\mathbf{x^{\prime }};\omega ),
\label{GF0} \\
G_b^{\mu \nu }(\mathbf{x},\mathbf{x^{\prime }};\omega ) &=&-b\left[ i\omega
\epsilon ^{\mu \nu 30}-\epsilon ^{\mu \nu 3i}\partial _{i}\right] f(\mathbf{x%
},\mathbf{x^{\prime }};\omega ),  \label{GFB} \\
G_{b^{2}}^{\mu \nu }(\mathbf{x},\mathbf{x^{\prime }};\omega ) &=&-{\tilde b}%
^{\mu }{\tilde b}^{\nu } f(\mathbf{x},\mathbf{x^{\prime }};\omega ),
\label{GFB2}
\end{eqnarray}%
whose evaluation requires the calculation on the action of the operators,
\begin{equation}
\left( k_0^{2}+\gv{\nabla} ^{2}\right)
=\left( k_0^{2}+\gv{\nabla} _{\bot }^{2}+\frac{\partial ^{2}}{\partial Z^{2}}\right),
\end{equation} 
and $\partial _i $ on the function $\frk{f}(\mbf{x},\mbf{x'};\omega)$, which is the far field approximation of $f(\mathbf{x}%
,\mathbf{x^{\prime }};\omega )$. Instead of applying these operators directly
on $\frk{f}(\mbf{x},\mbf{x'};\omega)$, we choose to act upon the exact expressions (%
\ref{FETA}) for $f_{\eta}(\mathbf{x},\mathbf{x^{\prime }};\omega )\;$ and
subsequently evaluate the results in terms of the stationary phase proposed previously.
The calculation is sketched in the Appendix \ref{APPA}, yielding the result
\begin{equation}
G^{\mu \nu }(\mbf{x},\mbf{x'}; \omega) =  \frac{1}{8\pi } \sum_{\eta=\pm 1} \frac{e^{ik_0  C_\eta (\theta)r}}{ g_\eta (\theta) } {H}_\eta{}^{\mu \nu  } (\mbf{\hat{n}})\frac{1}{r}\, e^{-ik_0 \mbf{\hat{n}}\cdot \mbf{x'} C_\eta (\theta)},
\label{FINALGF}
\end{equation}
with
\begin{equation}
{H}_\eta{}^{\mu\nu} (\mbf{\hat{n}}) = 
\begin{pmatrix}
1 & - i \eta |\tan\theta| \sin\phi & \ i \eta |\tan\theta|\cos\phi & 0 \\
 i \eta |\tan\theta| \sin\phi & -1 &\  i \eta |\sec\theta| & 0 \\
- i \eta | \tan\theta| \cos\phi  & \ - i \eta |\sec\theta| & -1 & 0\\
0 & 0 & \ 0 & -(1 +\eta  \beta |\sec\theta| ).
\end{pmatrix},
\label{MHGF}
\end{equation}
fulfilling the following symmetry property 
\beq
H_\eta{}^{\mu\nu}(-\omega)=(H_{-\eta}{}^{\mu\nu}(\omega))^*.
\label{SYMHMUNU}
\eeq
As another check of consistency, let us recover the conventional GF when $\beta=0$. Recalling
$C_\eta(\theta)=1$ and $g_\eta(\theta)=1$ in this case, the GF turns into  
\beq
G^{\mu \nu }(\mbf{x},\mbf{x'}; \omega) =  \frac{1}{8\pi }\frac{e^{ik_0 (r -\mbf{\hat{n}}\cdot \mbf{x'})}}{r} \sum_{\eta=\pm 1}  {H}_\eta{}^{\mu \nu  } (\mbf{\hat{n}})= \eta^{\mu\nu} \frac{1}{4\pi }\frac{e^{ik_0 (r -\mbf{\hat{n}}\cdot \mbf{x'})}}{r},
\label{GFB0}
\eeq  
since the non-diagonal terms in Eq. (\ref{MHGF}) cancel in the sum, while those in the  diagonal  get a factor of two.

\section{The electromagnetic fields in the radiation zone}

\label{EMFRAD}
In this section we calculate $\mbf{E}$ and $\mbf{B}$ in the far-field approximation, for 
 an arbitrary localized current $J_{\nu }(\mathbf{x^{\prime }},\omega)$.  We start from  the vector potential together with  the relation
\begin{equation}
A^{\mu }(\mathbf{x},\omega )=\frac{4\pi}{c}\int d^{3}x^{\prime }\;G^{\mu \nu }(\mathbf{x},%
\mathbf{x^{\prime }};\omega)\,J_{\nu }(\mathbf{x^{\prime }},\omega),
\label{AGEN}
\end{equation}
which, with the GF (\ref{FINALGF}) implemented, yields
\begin{equation}
A^{\mu}(\mbf{x},\omega )=\frac{1}{2c}\sum_{\eta= \pm 1}\frac{1}{g_\eta(\theta)}{H}_\eta{}^{\mu\nu} (\mbf{\hat{n}}) 
\frac{1}{r}e^{ik_0 C_\eta(\theta)r}
\int
d^{3}x^{\prime }e^{-i k_0 C_\eta (\theta)\hat{\mathbf{n}}\cdot \mathbf{x^{\prime }}}J_{\nu }(\mathbf{x^{\prime }},\omega),
\label{APART}
\end{equation}
where the only dependence of the GF on the source points $\mathbf{x}{%
^{\prime }}$ is in the exponential $
\exp (-i k_0 \hat{\mathbf{n}}\cdot \mathbf{%
x^{\prime }} \, C_\eta(\theta))$. Thus, the relevant integral is the space Fourier transform 
\begin{equation}
\mathcal{J}^{\nu }({}\mathbf{k}{}_{\eta },\,\omega )=\int
d^{3}x^{\prime }\,e^{-i k_0 C_\eta (\theta)\hat{\mathbf{n}}\cdot \mathbf{x^{\prime }}}\ J^{\nu }(\mathbf{x^{\prime }},\omega ),
\label{INT}
\end{equation}%
with 
\begin{equation}
\mathbf{k}{}_{\eta }=\mbf{\hat{r}} k_0 C_\eta(\theta),\qquad \mbf{\hat{r}}=(\sin \theta \cos \phi ,\,\sin \theta \sin \phi
,\,\cos \theta ),
\label{KETA1}
\end{equation}%
where $\theta$ and $\phi$ are the observation angles in spherical coordinates. Let us remark that no absolute values in the $\theta$-dependent angular functions appear in $\mbf{\hat{r}}$.   The  electromagnetic potential is then 
\begin{equation}
A^{\mu }(\mathbf{x},\omega )=\frac{1}{2c}\sum_{\eta=\pm 1 }\frac{1}{g_\eta(\theta)}{ H}_{\eta
}{}^{\mu \nu }(\hat{{\mbf r}})\; {\mathcal J}_{\nu }({}\mathbf{k}{}_{\eta },\,\omega
)\left( \frac{1}{r}e^{ik_0 C_\eta(\theta)r}\right).  \label{EMPOT1}
\end{equation}
Collecting the symmetry properties given in  Eqs. (\ref{SYMCG}) and (\ref{SYMHMUNU}) and using $J^\mu(\mbf{x}, -\omega)=(J^\mu(\mbf{x}, \omega))^*$, we arrive at the  relation
\beq
\qquad A^\mu_\eta(\mbf{x},-\omega)=(A^\mu_{-\eta}(\mbf{x},\omega))^*.
\label{SYMAMU}
\eeq
We can  verify again the correct limit $\beta=0$, since now  ${\mbf k}_\eta={\mbf{\hat n}k_0}$ independently of $\eta$, because $C_\eta(\theta)=1$, so that the sum in (\ref{EMPOT1}) reduces to ${\cal J}_\nu \sum_{\eta=\pm 1} H_\eta^{\mu\nu}$. As mentioned previously, the off-diagonal terms cancel and those in the diagonal yield $2 \eta^{\mu\nu}$. Remember also that $g_\eta(\theta)=1$ when $\beta=0$.

To calculate the electromagnetic fields $\mathbf{E}(\mathbf{x},\omega ) =i k_0 \mathbf{A}(\mathbf{x},\omega
)-\gv{\nabla} A^{0}(\mathbf{x},\omega )$ and $\mathbf{B}(\mathbf{x},\omega )
=\gv{\nabla} \times \mathbf{A}(\mathbf{x},\omega )$,
it is convenient to write the vector potential (\ref{EMPOT1}%
) in the form 
\beq
{A}^{\mu }=\sum_{\eta=\pm 1} A^\mu_\eta, \quad A^\mu_\eta= \mathcal{A}_{\eta }^{\mu }(\theta
,\phi )\left( \frac{1}{ g_\eta(\theta)}\frac{1}{2 c r}e^{i k_0 C_\eta(\theta) r }\right), \,\,\,\,\, 
\mathcal{A}_{\eta }^{\mu }(\theta ,\phi )={H}_{\eta
}{}^{\mu \nu }(\mbf{\hat{r}})\; {\mathcal J}_{\nu }({}\mathbf{k}{}_{\eta },\,\omega ).
\label{EMPOT2}
\eeq	
As usual, the form of the outgoing wave $\exp(ik_0 C_\eta(\theta)r )
/r$ in the
radiation zone provides an important simplification when considering the
action of the gradient operator. It is a direct calculation to show that
\begin{equation}
		{\gv \nabla}\Big( U(\theta, \phi) \frac{1}{r} e^{i k_0 C_\eta(\theta) r }\Big)=i k_0 {\mbf N}_\eta \Big( U(\theta, \phi) 
		\frac{1}{r} e^{i k_0 C_\eta(\theta) r }\Big),
		\label{GRADRAD}
\end{equation}
for an arbitrary angular function $U(\theta, \phi)$ in the far-field approximation, with
\beq
{\mbf N}_\eta=\Big[C_{\eta }(\theta)\,\hat{{\mbf r}}+\frac{\partial C_{\eta }(\theta)}{\partial \theta }\,
\gv{\hat{ \theta}}
\Big].
\label{DEFN}
\eeq
In accordance with the expression (\ref{GRADRAD}),  the equivalence
$\gv{\nabla}=ik_0 \mbf{N}_\eta$ holds
for the action of $\gv{\nabla}$ on the function between parentesis in (\ref{GRADRAD}), to first order in $1/r$. 
This is the generalization of  the familiar property, $\gv{\nabla} \rightarrow ik_0 \mbf{\hat r} $, in the scenario of conventional radiation.
Let us observe that the dependence on $\gv{\hat{ \theta}}$, which is not present in the standard case,   arises because the $\theta $-dependent phase factor $C_{\eta }(\theta)$ in the exponential $\exp(ik_0 C_\eta(\theta)r )$. The electromagnetic fields can also be decomposed into their $\eta$-contributions and they are 
\begin{eqnarray}
&&{\mbf B}=\sum_{\eta= \pm 1}{\mbf B}_\eta, \quad {\mbf B}_\eta=i k_0  \big(
{\mbf N}_\eta
 \times \,\mbf{ A}_\eta
 \big), \label{FINB} \\
&&{\mathbf E}=\sum_{\eta= \pm 1}{\mbf E}_\eta, \quad {\mbf E}_\eta=i k_0 \Big(\mbf{ A}_\eta-{\mbf N}_\eta \, {A}_{\eta }^{0}
\Big),\label{FINE}
\end{eqnarray}%
where we  recall that the expression of the vector potential $A^\mu$ in terms of the sources is given in Eq. (\ref{EMPOT2}). Equations (\ref{FINB}) and (\ref{FINE}) yield 
\beq
\mbf{N}_\eta \times \mbf{E}_\eta =ik_0 \big(\mbf{N}_\eta\times \mbf{A}_\eta\big)= \mbf{B}_\eta. \label{BNTE}
 \eeq
Then we obtain $ik_0 \big(\mbf{N}_\eta\times \mbf{E}_\eta\big)=\gv{\nabla} \times E_\eta =ik_0 \mbf{B}_\eta$ which readily implies Faraday law $\gv{\nabla}\times \mbf{E}(\mbf{x}, \omega)=ik_0 \mbf{B}(\mbf{x}, \omega)$.
 
The explicit  expression  for the electric field in spherical
 coordinates is
 \beq
\mbf{E}_\eta(\mbf{x},\omega) = i k_0 \left[ \left( A^r_\eta-C_\eta A^0_\eta \right) \hat{{\mbf r}} +\left( A^\theta_\eta-\frac{\partial C_\eta}{\partial \theta}A^0_\eta\right)\gv{\hat{\theta}} + A^\phi_\eta \,  \gv{\hat{\phi}} \right],
\label{ESPH}
\eeq 
such that  
\beq
\hat{{\mbf r}}  \cdot {\mbf E}_\eta=ik_0 \left( A^r_\eta-C_\eta A^0_\eta \right),
\label{RDOTE}
\eeq
which will be non-zero in general. 
{From Eqs. (\ref{FINE}) and (\ref{BNTE}),} we get
\beq
{\mbf E}\cdot {\mbf B}=\frac{ik_0}{2 c \,r}({\mbf E}_- \times {\mbf E}_+)\cdot \left(\mbf{ N}_- - \mbf{ N}_+\right),
\label{EDOTB}
\eeq
 showing that ${\mbf E}$
and  ${\mbf B}$ are {generally non orthogonal. Furthermore, from Eq. (\ref{BNTE}) and  (\ref{ESPH}),} we obtain the additional projection
\beq
\hat{{\mbf r}}  \cdot {\mbf B}_\eta=ik_0 \frac{\partial C_\eta(\theta)}{\partial \theta} A_\eta^\phi.
\label{RDOTB}
\eeq 
This, together with Eqs. (\ref{RDOTE}) and (\ref{EDOTB}) indicates that the triad $\hat{{\mbf r}}, \mbf{E}, \mbf{B} $ is not orthogonal as it is  in the conventional case.
The results (\ref{FINB}) and (\ref{FINE})  completely describe the properties of the radiation
emitted by an arbitrary current $J^{\mu }(\mathbf{x}{^{\prime }},\omega )$
in a chiral material having ${\tilde b}_{\mu }=(0,0,0,b)$.
 
The ordinary properties of the radiation field when $\beta=0$ are easily recovered from previous equations. In this case 
$\mbf{ {N}_+}=\mbf{ {N}_-}=\hat{{\mbf r}} $,  with $ C_\eta(\theta)=1$, which yield ${\mbf E}\cdot {\mbf B}=0$ and  $\hat{{\mbf r}}\cdot {\mbf B}=0$. Also, $A^\mu_+({\mbf x}, \omega)= A^\mu_- ({\mbf x}, \omega)= \frac{e^{ik_0 r}}{2 c\, r} \, {\cal J}^\mu(\gv{\hat{r}}k_0, \omega)$, which allows to write Eq. (\ref{RDOTE}) as
\beq
\hat{{\mbf r}}  \cdot {\mbf E}=ik_0 \left( \hat{{\mbf r}}\cdot \mbf{A}-A^0 \right)=i\frac{e^{ik_0 r}}{2 c\, r} (k_0 \, \hat{{\mbf r}} \cdot \gv{\cal J}-k_0 {\cal J}^0 )=-i\frac{e^{ik_0 r}}{2 c\, r} K_\mu {\cal J}^\mu(\gv{\hat{r}}k_0, \omega)=0,
\label{RDOTECONV}
\eeq
where the last term is zero by current conservation, with $ K_\mu=(k_0, k_0 \hat{{\mbf r}})$ in the radiation regime.

The explicit form of the electromagnetic potential in terms of an arbitrary current ${\cal J}^\mu({\mbf k}_\eta, \omega)$,  together with the components of the electromagnetic fields 
and a general expression for the spectral distribution of the  {radiation, are given} in the  Appendix \ref{APPB}.

 Until now we have only considered the  electromagnetic radiation produced in  an ideal chiral medium  with refraction index $n=1$. This case corresponds to what is called vacuum Cherenkov radiation in the literature \cite{POTTING1,POTTING2}, {where it} is assumed that the standard vacuum is filled with background fields codifying LIV, whose electromagnetic effects turn out to be  analogous to a material medium. On the other hand, non-magnetic ($\mu=1 $) chiral materials have refraction indices $n >1$, which we need to take into account. The transformations relating both regimes are presented in the Appendix {\ref{APPD}} and for our immediate purposes they include the following replacements
\beq
q \,\, \rightarrow \,\, q/n, \quad c  \,\, \rightarrow \,\, c/n, \qquad b  \,\, \rightarrow \,\, b/n, \qquad \beta \,\, \rightarrow \,\, {\tilde \beta}= \frac{c b}{\omega}\frac{1}{n^2}= { \beta} \frac{1}{n^2}, 
\label{REPLACE}
\eeq
In the following we denote  with a tilde the quantities which now are presented for arbitrary $n$ mainly  to distinguish them from those in previous notation with $n=1$. In an abuse of notation we do not make this distinction in the resulting  electromagnetic potentials and fields.

\section{The Cherenkov radiation}
\label{CHRAD}

Now we apply the general method developed in the previous sections to the case of a charge $q$ moving in chiral matter with constant
velocity, ${{\gv v}}= v {\gv {\hat e}}_z$, parallel to ${\gv b}$, along the $z$-axis, in order to
maintain axial symmetry. The developments in this section rely heavily on {previous} results {together of those in} the Appendixes \ref{APPB} and \ref{APPC}, {all of which were obtained } for $n=1$. Nevertheless, as  {necessary} for the description of non-magnetic chiral matter with  refraction index $n >1 $, we { need to} perform the   substitutions indicated in Eq. (\ref{REPLACE}). {These are carried over for all  the quantities  we use  for the remaining  calculations and plots, and are indicated by adding an upper tilde over the respective seed function previously calculated for $n=1$. In other words, ${\tilde F}(n)$ is obtained from $F(n=1)$ after making the substitutions indicated in Eq. (\ref{REPLACE}).  The remaining symbols $q,c,b, \omega, \beta$ retain their original meaning indicated in the preceding sections. }

\subsection{The electromagnetic fields}

\label{EMFCH}

The sources in the frequency space  are 
\begin{eqnarray}
\rho (\mathbf{x}^{\prime},\omega) = \frac{q}{nv} \delta(x^{\prime})
\delta(y^{\prime})e^{i \omega \frac{z^{\prime}}{v}}, \qquad \mathbf{J}(%
\mathbf{x}^{\prime},\omega) = \frac{q}{n} \delta(x^{\prime}) \delta(y^{\prime})e^{i
\omega \frac{z^{\prime}}{v}}. \label{SOURCESCH}
\end{eqnarray}
In order to have a well defined limiting process in our calculation, we
follow Refs. \cite{Panofsky,PRDOJF} integrating the charge
trajectory in the interval $z\in (-\xi ,\xi )$ and taking the limit $\xi
\rightarrow \infty $, at the end of the calculation. From the charge and current densities,
we obtain 
\ba
\mathcal{J}^0 ({\mathbf{\tilde{ k}}_\eta}, \omega)=2\frac{q}{n}\frac{c}{v}\frac{\sin [\xi \, \tilde{\Xi} _{\eta }]}{\tilde{\Xi}
_{\eta }}, \qquad  \mathcal{J}^3 ({\mathbf{\tilde{k}}_\eta}, \omega)=2\frac{q}{n}\frac{\sin [\xi \, \tilde{\Xi} _{\eta }]}{\tilde{\Xi} _{\eta }}, \label{SOURCESFT}
\ea
where now we have
\ba
&& \tilde{\Xi}_\eta(\omega, \theta)= \frac{\omega }{v }\Big(1-\frac{ n v }{c} \tilde{C}_{\eta }(\theta )\cos \theta \Big), \quad \tilde{C}_\eta(\omega, \theta)=
\sin^2\theta+ \cos^2\theta \sqrt{1+ \eta {\tilde \beta} \sec\theta} \label{TILDECHI},\\
&& \tilde{g}_\eta (\omega, \theta)=\frac{1}{\left( 1+\eta {\tilde \beta} \sec \theta \right)^{1/4}}
\sqrt{ \left(1+ \eta {\tilde \beta} \sec\theta\right) \left(1 +\frac{\eta {\tilde \beta}}{2} \sec\theta\right) + \frac{{\tilde \beta}^2}{4}{\tan^2\theta}},
\label{TILDEG}
\ea
and $\mathbf{\tilde{k}}{}_{\eta }=\mbf{\hat{r}} \frac{n \omega}{c}  {\tilde C}_\eta(\theta) $, {with $\mbf{\hat{r}}$ being} the unit vector in the direction of the observation point. Using Eq. (\ref{MHGF}), the components of the electromagnetic potential (\ref{EMPOT1}) are conveniently written as 
\begin{eqnarray}
A_{\eta }^{0} &=&\frac{1}{v}\tilde{Q}_\eta, \qquad  A_{\eta }^{z} =\frac{1}{c}\left( 1+\eta {\tilde \beta} |\sec \theta| \right) \tilde{Q}_\eta, \label{ACONV1}\\
A_{\eta }^{x} &=&\frac{1}{v} \Big( i\eta \sin \phi |\tan \theta| \Big) \tilde{Q}_\eta, \qquad 
A_{\eta }^{y} =-\frac{1}{v} \Big(i\eta \cos \phi |\tan \theta| \Big) \tilde{Q}_\eta, \label{ACONV2}
\end{eqnarray} with 
\ba
\tilde{Q}_\eta &=&\frac{q}{n}\frac{%
1 }{{\tilde g}_{\eta }(\theta )}\,\, \frac{\sin [\xi \,
\tilde{\Xi} _{\eta }]}{\tilde{\Xi} _{\eta }} \frac{1}{r} 
e^{i \frac{n\omega}{c}  {\tilde{C}}_{\eta }(\theta)r} .
\label{QETA}
\end{eqnarray}%
An  important simplifying feature in the calculation is the delta-like behavior of the function $\tilde{Q}_\eta$ in the limit 
\beq
\lim_{\xi \rightarrow \infty} \frac{\sin[\xi \tilde{\Xi}_\eta ]}{\tilde{\Xi}_\eta}  = \pi \delta \left( \frac{\omega}{v}(1-\frac{nv}{c}\cos\theta \, \tilde{C}_\eta (\theta))\right),\label{ANGLECOND1}
\eeq
which determines the allowed Cherenkov angles, 
\beq
 \cos \theta \, \tilde{C}_\eta(\theta)=\frac{c}{ n v},
 \label{ANGLECOND2}
\eeq
in complete analogy with the ordinary case. {Since} we  discard the imaginary contributions to  $\tilde{C}_\eta(\theta)$ in  Eq. (\ref{TILDECHI}), the allowed values are positive and we conclude that the resulting angles $\theta_{\eta}$, determined from Eq. (\ref{ANGLECOND2}), are in the range $[0, \pi/2]$. In other words, there is radiation only in the forward direction.

An additional advantage is that now we can replace $|\cos \theta|$ by $\cos \theta$ in all previous functions. 
Making explicit the\ axial symmetry in polar coordinates, we introduce ${\gv {\hat \phi}}=\left[ -\sin
\phi \;{\gv {\hat e}}_x+\cos \phi \,  {\gv {\hat e}}_y \right]$, leading to 
\begin{equation}
\mathbf{A}_\eta(\mathbf{x},\omega) =  A^z_\eta(\mathbf{x},\omega)
\cos\theta \, \hat{{\mbf r}} - A^z_\eta(\mathbf{x},\omega) \sin\theta \, {\gv {\hat \theta}}+ A^\phi_\eta(\mathbf{x},\omega) {\gv { \hat \phi}},
\label{EMPOTESF}
\end{equation}
in spherical coordinates, with
\begin{equation}
A^\phi_\eta(\mathbf{x},\omega) = -i\eta   \frac{1}{v}\tan\theta \, 
{\tilde Q}_\eta.  \label{DEFAPHI}
\end{equation}
The associated electromagnetic fields are 
\ba
&& \hspace{-0.8cm}\mathbf{E}(\mathbf{x},\omega) = \sum_{\eta=\pm 1} \mathbf{E}_\eta(\mathbf{x},\omega), \label{TEF}\\
&& \hspace{-0.8cm} \mathbf{E}_\eta(\mathbf{x},\omega) = i \frac{n \omega }{c}  \left[ \left( \cos \theta
A^z_\eta-\tilde{C}_\eta(\theta) A^0_\eta\right) \mbf{\hat{r}} -\left( \sin\theta A^z_\eta+\frac{%
\partial \tilde{C}_\eta(\theta)}{\partial \theta}A^0_\eta\right)\gv{\hat{\theta}} + A^\phi_\eta \, \gv{\hat{%
\phi}} \right],  \label{ETAEMF} \\
&&  \hspace{-0.8cm} \mathbf{B}(\mathbf{x},\omega) = \sum_{\eta=\pm 1} \Big(
\tilde{C}_{\eta }(\theta)\,{\mathbf{\hat{r}}}+\frac{\partial \tilde{C}_{\eta
}(\theta)} {\partial \theta } 
\hat{\gv \theta}
\Big)\times \mathbf{E}_\eta(\mathbf{x},\omega),\label{TBF} \\
&& \hspace{-0.8cm} \mathbf{B}(\mathbf{x},\omega) = i \frac{n \omega }{c} \sum_{\eta= \pm 1} \left[\frac{\partial \tilde{C}_\eta(\theta)}{\partial \theta}A^\phi_\eta \, \mbf{\hat{r}}
-\tilde{C}_\eta(\theta)  A^\phi_\eta \, \gv{\hat{\theta}}- \left(\tilde{C}_\eta(\theta)\sin\theta + \frac{%
\partial \tilde{C}_\eta(\theta)}{\partial\theta}\cos\theta \right) A^z_\eta \, \gv{\hat{\phi}}\right].
\label{TBEXPL}
\ea

\subsection{The spectral distribution of the radiated energy}

\label{SPECTRALCH}

From Eq. (\ref{MODUS}) we recover the standard energy-momentum conservation law with the usual Poynting vector, $\mathbf{S} = \frac{c}{4\pi}\mathbf{E}\times \mathbf{B}$, when we adopt $b_0=0$. Taking into account the fields (\ref{TEF}-\ref{TBEXPL}), the spectral energy distribution (SED)
\begin{eqnarray}
 && 
\frac{d^2E}{d\Omega d\omega}= \frac{c}{4\pi^2} r^2 \, \hat{\mathbf{r}} \cdot {\rm Re} \left[
\mathbf{E}^*(\mathbf{x},\omega) \times \mathbf{B}(\mathbf{x},\omega)\right],
\label{SPECDIST0}
\end{eqnarray}
reads
\begin{equation}
\begin{split}
\frac{d^2E}{d\Omega d\omega} & = \frac{n \omega^2q^2}{4\pi^2 c^3}\left[ \,  \sum_{\eta=\pm 1} \frac{%
\sin^2[\xi \, {\tilde \Xi}_\eta ]}{{ \tilde \Xi}^2_\eta}
\mathcal{\tilde {T}}_{1,\eta}(\omega,\theta) \frac{1}{{\tilde g}^2_\eta(\theta)} \right. \\ &  \left.
+\frac{\sin[\xi \, \tilde{\Xi}_+ ]}{\tilde{\Xi}_+} \frac{\sin[\xi \, \tilde{\Xi}_- ]}{\tilde{\Xi}_-} \left[
\mathcal{\tilde{T}}_{2,+}(\omega,\theta) +\mathcal{\tilde{T}}_{2,-}(\omega,\theta)\right] \frac{%
\cos(\frac{n\omega}{c}  (\tilde{C}_+(\theta)-\tilde{C}_-(\theta))r)}{\tilde{g}_+(\theta)\tilde{g}_-(\theta)}\right].
\end{split}%
\label{SPECDIST}
\end{equation}

To express  ${\cal \tilde{T}}_{1,\eta}$ and  ${\cal \tilde{ T}}_{2,\eta}$ in a compact way, it is first convenient to introduce the auxiliary functions
\ba
&& \tilde{p}_\eta(\omega, \theta)= \sin\theta + \eta {\tilde \beta} \tan\theta
+ \frac{c}{n v}\frac{\partial \tilde{C}_\eta}{\partial\theta} , \qquad \tilde{q}_\eta(\omega, \theta)= \cos\theta+ \eta {\tilde \beta}  -\frac{c}{n v}\tilde{C}_\eta, 
\label{PQETA} 
\ea
which yield
\ba
&& \mathcal{\tilde{T}}_{ 1,\eta }  =\left( \tilde{p}^2_\eta + \frac{c^2}{n^2 v^2}\tan^2\theta \right)\tilde{C}_\eta 
+ \tilde{p}_\eta \tilde{q}_{\eta}\frac{%
\partial \tilde{C}_\eta}{\partial\theta},\quad 
\mathcal{\tilde{T}}_{2,\eta}  =\left( \tilde{p}_- \tilde{p}_+ - 
\frac{c^2}{n^2 v^2}\tan^2\theta \right)\tilde{C}_{-\eta} +\tilde{p}_\eta \tilde{q}_{-\eta} \frac{%
\partial \tilde{C}_{-\eta}}{\partial\theta}.\nonumber \\
\label{CALTETA} 
\ea
For simplicity we have not written  the dependence upon $(\omega, \theta)$ in Eqs. (\ref{CALTETA}). In the Appendix  \ref{APPB} we list the symmetry properties of some of the above functions under the change $\omega \rightarrow - \omega.$ From the condition (\ref{ANGLECOND2}), we note that  $\tilde{C}_{+}=\tilde{C}_-$ only when $b=0$, so that the resulting angles $\theta_{\pm}$ will be different in a chiral media ($b\neq0$),  only coinciding when the material is nonchiral. This, together with the delta-like limit in Eq. (\ref{ANGLECOND1}), means that the cross {term} in Eq. (\ref{SPECDIST}),
\begin{equation}
\frac{\sin[\xi \, \tilde{\Xi}_+ ]}{\tilde{\Xi}_+} \frac{\sin[\xi \, \tilde{\Xi}_- ]}{\tilde{\Xi}_-},
\end{equation}
becomes zero in the final limit $\xi \rightarrow \infty $, leading to the following simpler expression for the SED
\begin{equation}
\frac{d^2E}{d\Omega d\omega}  = \frac{n \omega^2q^2}{4\pi^2 c^3} \,  \sum_{\eta=\pm 1} \frac{%
\sin^2[\xi \, {\tilde \Xi}_\eta ]}{{ \tilde \Xi}^2_\eta}\mathcal{\tilde{T}}_{1,\eta}(\omega,\theta) \frac{1}{{\tilde g}^2_\eta(\theta)}. 
\label{SPECDIST1}
\end{equation}

\subsection{Determination of the Cherenkov angles}

\label{CHANGLES}

\begin{figure}[h!]
\centering
\includegraphics[scale=0.75]{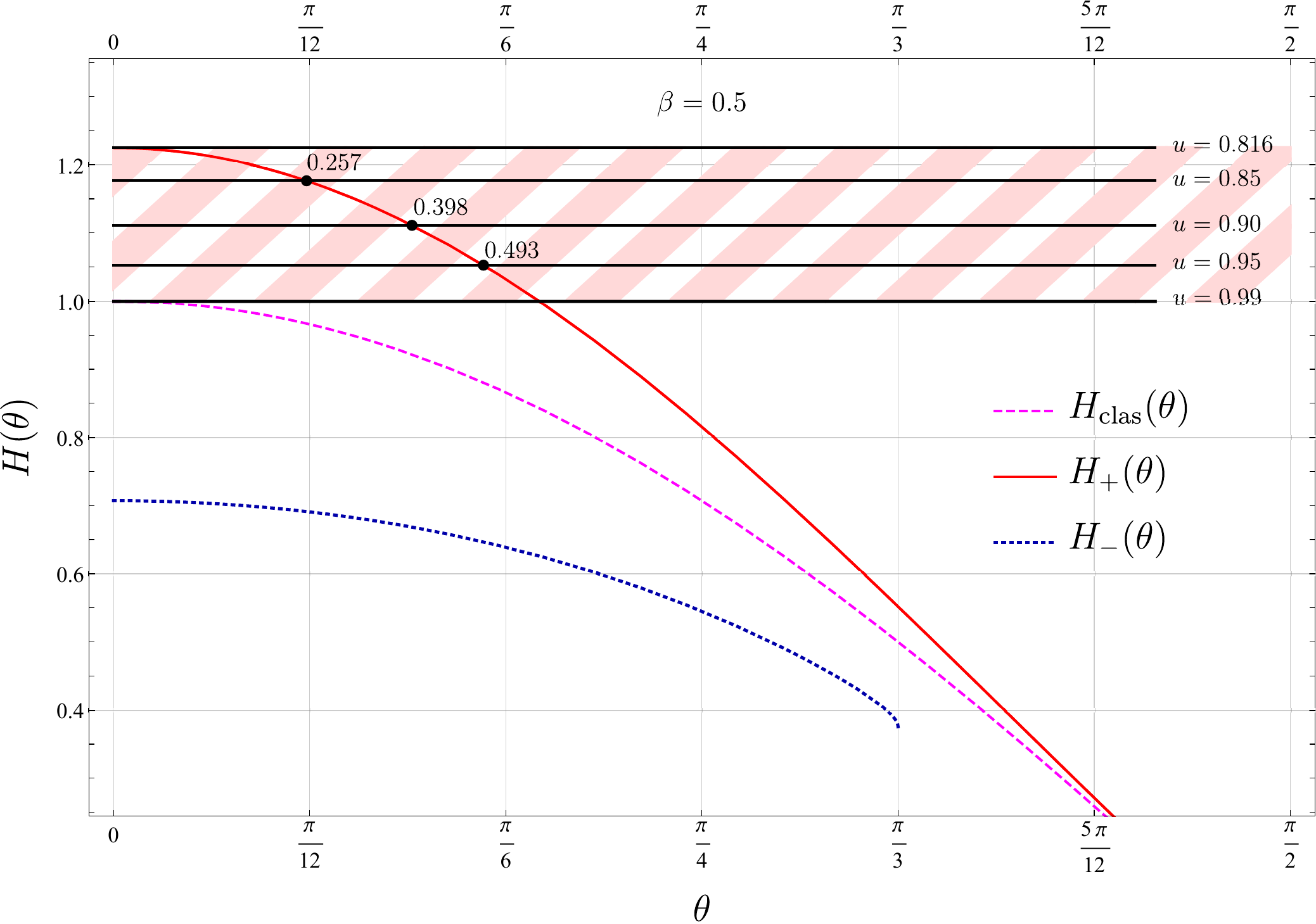}
\caption{Plot of the functions  $H_+(\theta)$ (solid line, red),  $H_{\rm clas}(\theta)$ (dashed line, magenta)   and $H_-(\theta)$ (dashed line, blue), for $\beta=bc/\omega=0.5$ \,   ($n=1)$. The horizontal lines  labeled by $u=v/c$ correspond to $H=1/u$  in the ordinate. The Cherenkov angles  are in radians.  Even though $n=1$ the hatched region indicates the presence of CHR in chiral matter.}
\label{FIG1} 
\end{figure}

Now we consider in detail the Cherenkov condition (\ref{ANGLECOND2}), which can be expressed in terms of the function 
\beq
H_\eta(\theta)\equiv \cos\theta\left(\sin^2\theta+ \cos^2\theta \sqrt{1+ \eta \, {\tilde \beta} \sec\theta}\right), \label{HTHETA}
\eeq
in such a way the Cherenkov angles, $\theta_\eta$ (with $ \eta=+, -, \rm clas$), are determined by the intersection of the $H_\eta(\theta)$ curves with the horizontal lines $c/(nv)$, in accordance with the relation (\ref{ANGLECOND2}). See the plot of $H_\eta(\theta)$ intersecting the horizontal lines in Fig. \ref{FIG1}, for $n=1$.
In the limit of standard electrodynamics in vacuum ($n=1, b=0$), one has $H_{\rm clas}=\cos \theta$, implying the known condition $\cos \theta=c/v $, which forbids the Cherenkov radiation.

\begin{figure}[h!]
\centering
\includegraphics[scale=0.75]{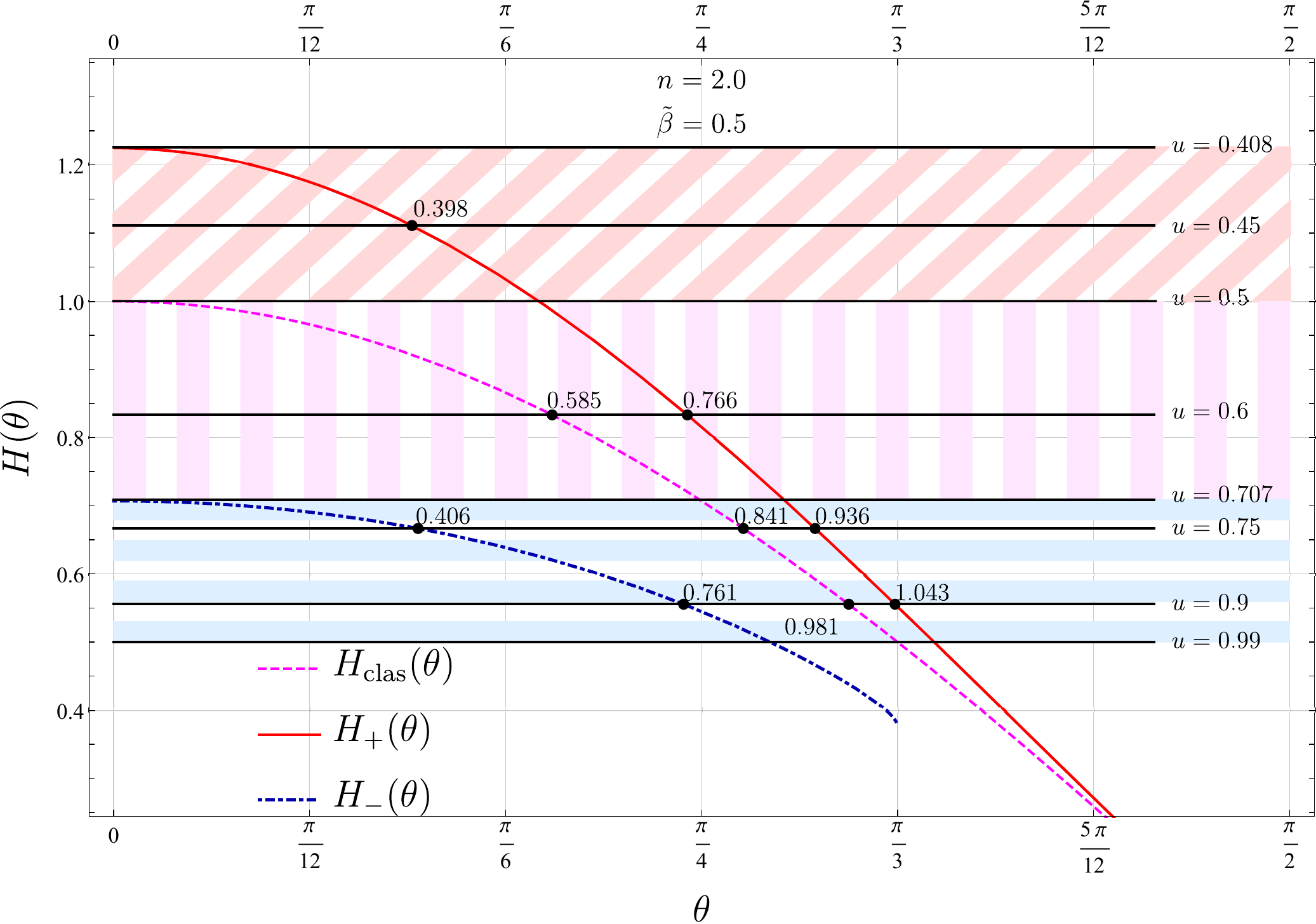}
\caption{Plot of the functions  $H_+(\theta)$ (solid line, red), $H_{\rm clas}(\theta)$ (dashed line, magenta) and $H_-(\theta)$ dash-dotted line, blue), for ${\tilde \beta}=0.5, \, n=2$. The horizontal lines  labeled by $u=v/c$ correspond to $H=1/(nu)$  in the ordinate.  The Cherenkov angles  are in radians. Region A: upper white area, Region B: oblique hatched area (orange) , Region C: vertical hatched area (light magenta) and Region D: horizontal hatched area (light blue.) }
\label{FIG2}
\end{figure}

In the following we list some  important properties of the functions $H_\eta(\theta)$. They are decreasing functions of $\theta$ that satisfy $H_- < H_{\rm clas} < H_+$, for all values of $\theta$. This justifies the name of outer (inner) cone that we give to the  radiation emitted at the angles $\theta_+$ ($\theta_-$), respectively.
 
The function $H_+$ is real for any value of ${\tilde \beta}$, while $H_-$ has the following restrictions: (i) it is imaginary when ${\tilde \beta} >1$, so that it does not contribute to the radiation in this case;  (ii) when  ${\tilde \beta} < 1$, it is real only in the interval $0 < \theta < \arccos {\tilde \beta}$, which constrains its contribution. The maximum values of each function at the origin are
\beq
H_+(\theta=0)=\sqrt{1+{\tilde \beta}}, \quad H_{\rm clas} (\theta=0)=1, \quad H_-(\theta=0)=\sqrt{1-{\tilde \beta}}.
\label{MAXH}
\eeq
For a given material (fixed $n > 1$), the evolution  of the Cherenkov angles $\theta_+, \theta_-, \theta_{\rm clas}, $ as a function of the particle velocity, $u=v/c$, are described as follows (the classical case is incorporated for comparison):
\begin{itemize}
\item {\bf Region A: }  No Cherenkov radiation when 
\beq
\sqrt{1+{\tilde \beta}}< 1/(nu),\label{COND1}
\eeq
corresponding to the upper white area above the $u=0.408$ horizontal line in Fig. \ref{FIG2}.
\item {\bf Region B:} The region in which only $\theta_+ $ arises is in the interval 
\beq
1< 1/(nu) < \sqrt{1+{\tilde \beta}}, \label{COND2}
\eeq
defined by the orange oblique hatched area in Fig. \ref{FIG2}. This is the vacuum Cherenkov radiation in the LIV case $(b\neq 0, n=1)$, which is not allowed in the classical situation \cite{POTTING1,POTTING2}.

\item  {\bf Region C: } The region in which both $\theta_+$ and $\theta_{\rm clas}$ are present is in the interval   
\beq
\sqrt{1-{\tilde \beta}} < 1/(nu) < 1, \label{COND3}
\eeq
shown by the pale magenta vertical hatched area in Fig. \ref{FIG2}.
\item  {\bf Region D: } The area in which all three angles, $\theta_+$, $\theta_{\rm clas}$ and $\theta_-$ coexist is in the interval  
\beq
{\tilde \beta}\left( 1- {\tilde \beta}^2\right)  < 1/(nu) < \sqrt{1-{\tilde \beta}}, \label{COND4}
\eeq
marked as the pale blue horizontal hatched area in Fig. \ref{FIG2}.
\item {\bf Region E: } The lower region in which only $\theta_+$ and $\theta_{\rm clas}$ arise corresponds to the interval   
\beq
1/n < 1/(nu) < {\tilde \beta}\left( 1- {\tilde \beta}^2\right). \label{COND5}
\eeq
\item The lower limit for $H=1/(nu)$ is $H=1/n$ and is given by the maximum  velocity  $u=1$.
\end{itemize}
The case of vacuum {Cherenkov} radiation  ($n=1$) is illustrated in the Fig. \ref{FIG1} for $\beta=0.5$. 
A generic case for $n=2$ and 
${\tilde \beta}=0.5$ is shown in the Fig. \ref{FIG2}. Regions A, B, C and D are indicated in the figure caption. In this case the lower
limit for $H$ is $1/2$, as approximately indicated by the horizontal line labeled by $u=0.99$. The values of the Cherenkov angles for different velocities corresponding to this case are also shown in Table \ref{TABLE1}. From the left panel of Fig.\ \ref{FIG01}, we appreciate that the Cherenkov angles obtained in this case, with 
our choice for  the SPA in Eq.\ (\ref{APPK}),
fall in a region where the classical SPA approximates very well the exact numerical values shown in the figure.
  
\begin{table}[ht]
\begin{center}
\begin{tabular}{cccc}
\hline 
\hline
{ $ u $ } \qquad   & { $ \theta_+(\rm rad) $ } \qquad & { $ \theta_{\rm clas}(\rm rad) $ } \qquad  & { $ \theta_-(\rm rad) $ } \qquad \tabularnewline
\hline 
\hline
$0.45$ \qquad & \qquad $0.398$ \quad & \qquad $0.0$ \quad &\qquad $0.0$ \quad \tabularnewline
$0.60$ \qquad & \qquad $0.766$ \quad & \qquad $0.585$ \quad &\qquad $0.0$ \quad \tabularnewline
$0.75$ \qquad & \qquad $0.936$ \quad & \qquad $0.841$ \quad &\qquad $0.406$ \quad \tabularnewline
$0.90$\qquad & \qquad $1.043$  \quad & \qquad $0.981$ \quad &\qquad $0.761$ \quad  \tabularnewline
\hline
\hline 
\end{tabular}
\caption{The possible Cherenkov angles for ${\tilde \beta}=0.5$ and $n=2$, exhibited in Fig. \ref{FIG2}.}
\par
\label{TABLE1}
\end{center}
\end{table}

\subsection{Cherenkov angles $\theta_+$, $\theta_-$ as a function on $\beta$}
\label{CHANGALPHA}

\begin{figure}[h!]
\centering
\includegraphics[width=0.45 \textwidth]{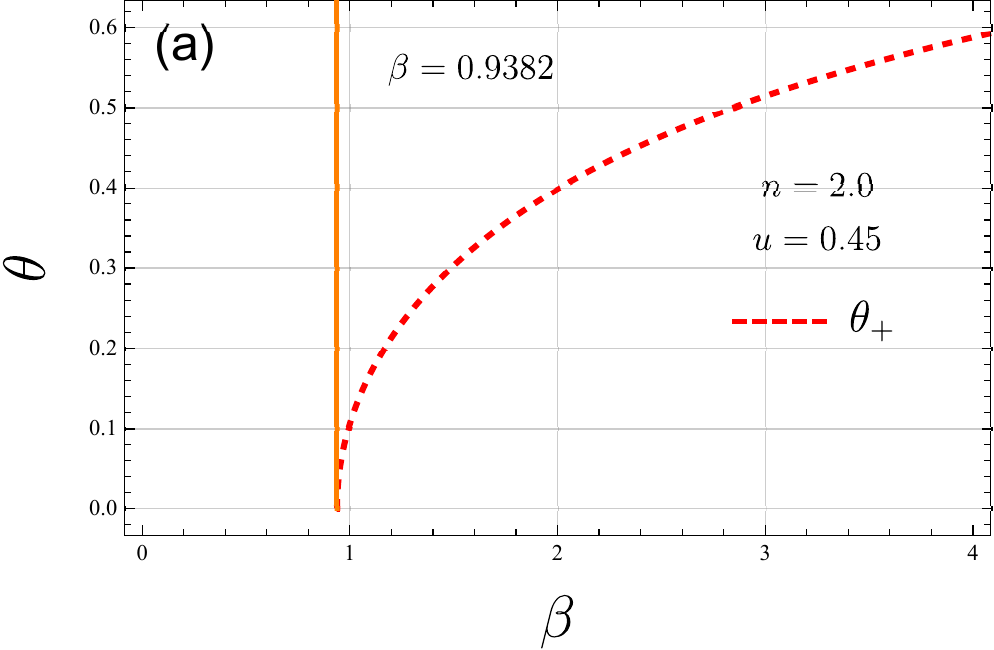}
\hspace{.5cm}
\includegraphics[width=0.45 \textwidth]{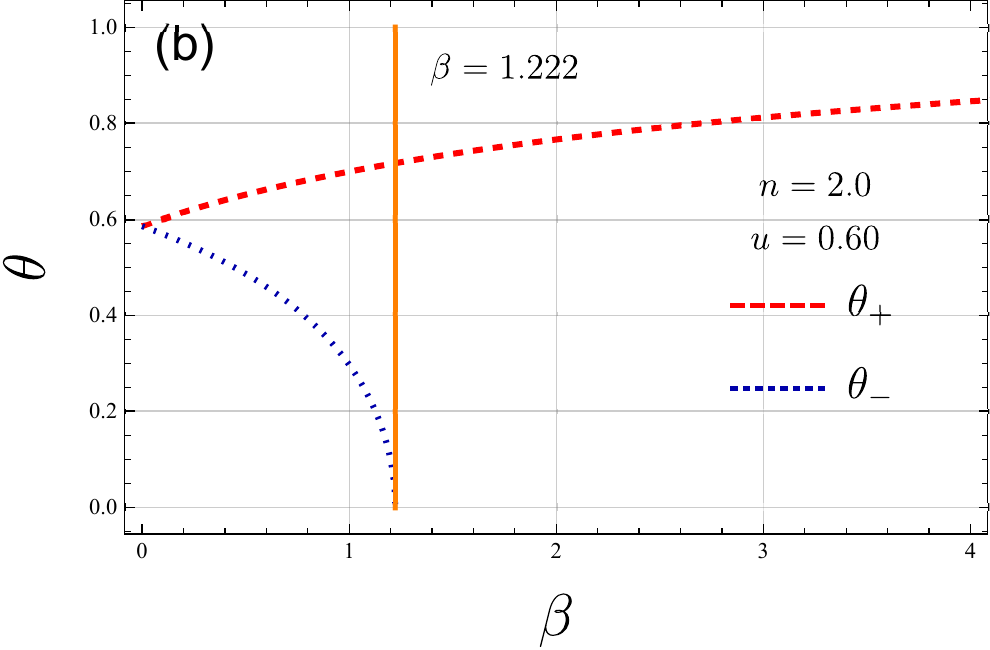}
\\
\vspace{0.5cm}
\includegraphics[width=0.45 \textwidth]{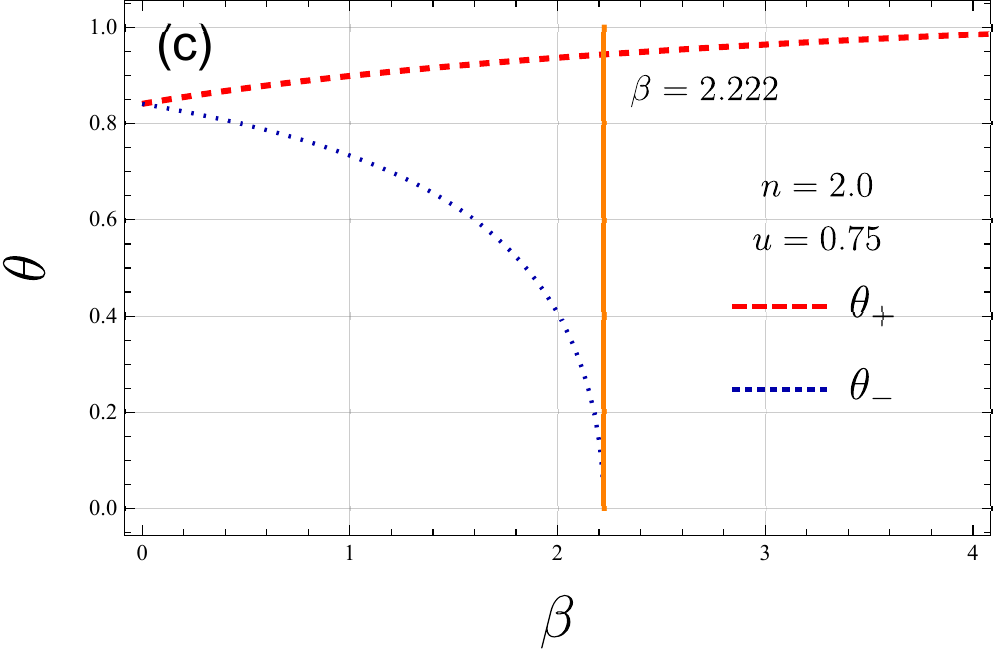}
\hspace{.5cm}
\includegraphics[width=0.45 \textwidth]{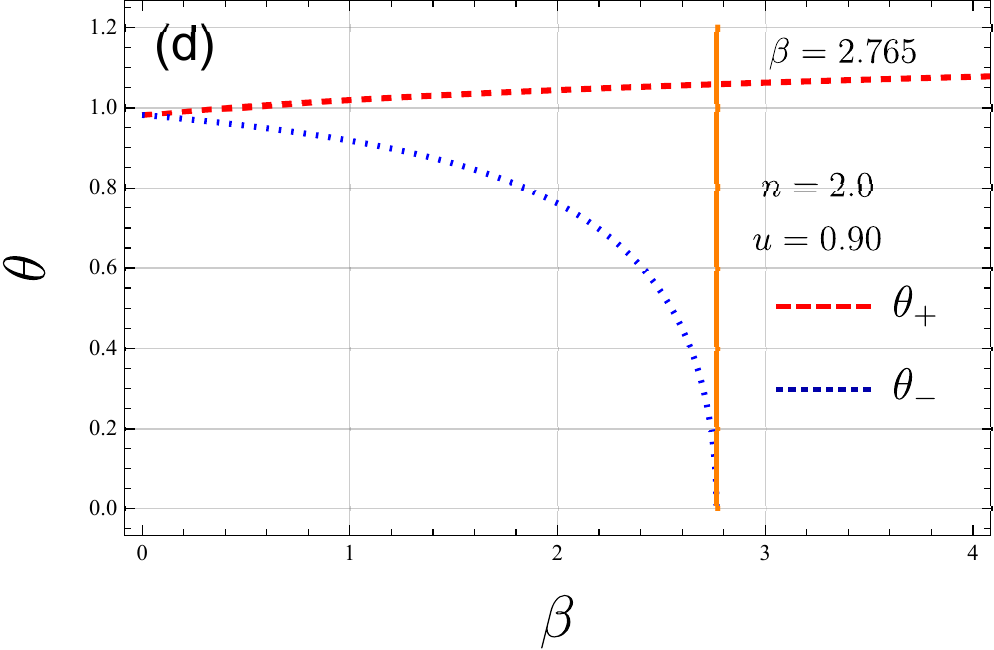}
\caption{The Cherenkov angles $\theta_+, \, \theta_-$,  for $n=2$ and different values of $u$,  as a function of $\beta$. Panel (a): $\theta_+$ vanishes at $\beta=\beta^C=0.938$. Panel (b): $\theta_-$ vanishes at $\beta=\beta^{C}=1.222$.  Panel (c) :  $\theta_-$ vanishes at $\beta=\beta^C=2.222$. Panel (d): $\theta_-$ vanishes at $\beta=\beta^C=2.765$. }
\label{FIG25}
\end{figure}

In this subsection we investigate the dependence of the Cherenkov angles upon the chiral parameter $\beta=bc/\omega$, as a function of $n$ and $u$, paying attention to the points where these angles are cut down, as shown in the Fig. \ref{FIG25}. In other words, we are interested in the functions $\theta_\pm(\beta)$ for fixed $n$ and $u$. Let us recall that $\beta(n)= \beta/n^2$. It can be shown that $\theta_+(\beta)$ ($\theta_-(\beta)$) is an increasing (decreasing) function of $\beta$.

 We distinguish two cases: 
\begin{enumerate}
    \item $nu < 1$. From Eqs. (\ref{COND2}) and (\ref{COND3})  we have only $\theta_+$. The angle  $\theta_+$ starts at $\beta^{\text{C}}$, where $\theta_+(\beta^{\text{C}}) = 0$. By substituting this in Eq. (\ref{HTHETA}), we obtain 
    \begin{equation}
        \beta^{\text{C}} =\frac{1}{u^2}- n^2,
    \label{NUMENUNO}\end{equation}
    as we can see from Fig. \ref{FIG25} (a).
    
    \item $nu > 1$. In this case we can have both $\theta_+$ and $\theta_-$,  which start at $\beta=0$. Since $\theta_-$ is a decreasing function of $\beta$, we determine a cutoff $\beta^{\text{C}}$ by setting  $\theta_-(\beta^{\text{C}}) = 0$. Replacing $\theta_-(\beta^{\text{C}}) = 0$ in Eq. (\ref{HTHETA}) yields
    \begin{equation}
        \beta^{\text{C}} = n^2-\frac{1}{u^2},
    \label{NUMAYUNO}\end{equation}
    which can be verified in  Figs. \ref{FIG25}(b-d).    
\end{enumerate}

\begin{figure}[h!]
\centering
\includegraphics[width=0.45 \textwidth]{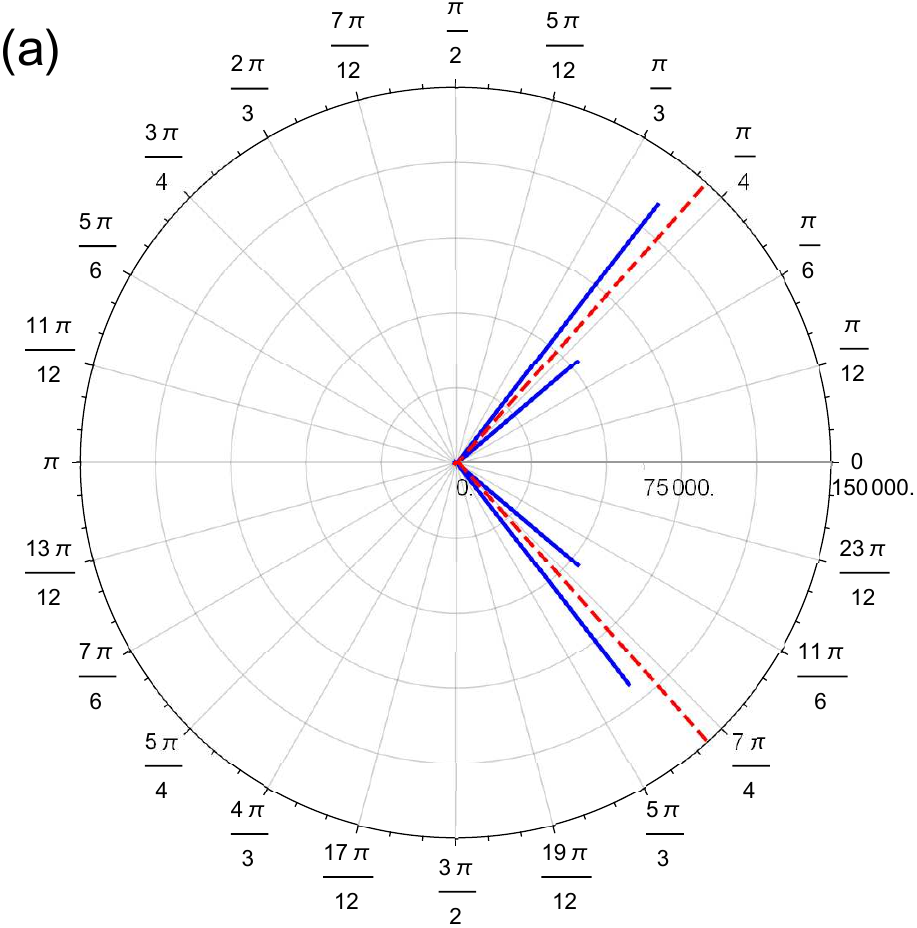}
\hspace{.5cm}
\includegraphics[width=0.45 \textwidth]{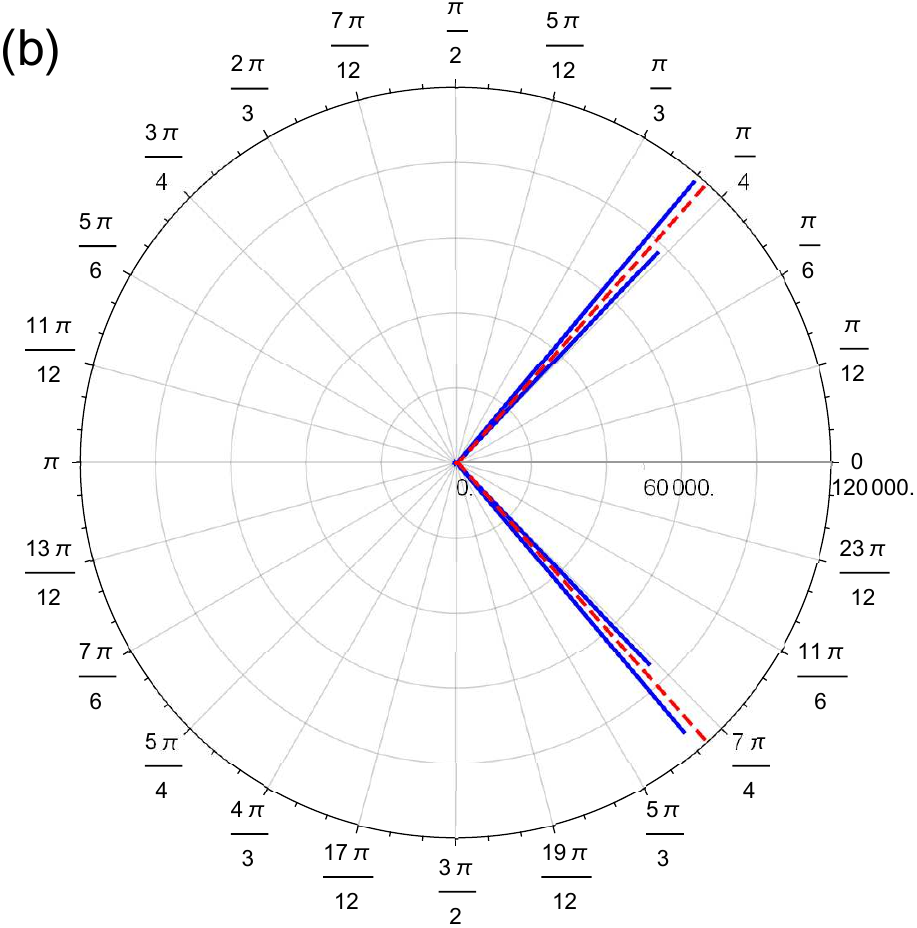}
\caption{The angular distribution   for the radiated energy per unit frequency  (solid blue line) in a chiral dielectric medium with $n=2$, $\omega/c =14 \;\text{$\mu$m}^{-1}$, $u=0.75$ and $\xi=100\;\text{$\mu$m}$. Panel (a) : ${\tilde \beta}=0.3$,  and Panel (b) :  ${\tilde \beta}=0.1$.   The dashed red line corresponds to the conventional  Cherenkov cone.  The charge moves from left to right.}
\label{FIG3}
\end{figure}

To close this section we present the 
figures \ref{FIG3}(a-b)  showing the SED for  $n=2$,  and the choices $\beta = 1.2$ and $\beta = 0.4$, which yield ${\tilde \beta}=0.3$ and ${\tilde \beta} =0.1$,  respectively.
In both figures the solid blue line corresponds to $\xi=100 \;\text{$\mu$m}$  with $u=0.75$ and  $\omega/c =14 \; \text{$\mu$m}^{-1}$. 
We notice that as long as  $\beta$ takes lesser values, the splitting between $\theta_+$ and $\theta_-$ closes up and the difference in the amplitudes of the lobes decreases. 
In Fig. \ref{FIG3}a the angles of the chiral Cherenkov cones are $\theta_+=0.908$ and $\theta_- =0.700$, and in Fig. \ref{FIG3}b they are $\theta_+=0.868$ and $\theta_-=0.807$. All plotted SED  are expressed in units of the common factor $q^2/(4 \pi^2 c^3)$ . 

\section{Total radiated energy per unit frequency}
\label{TOTRADEN}

In this section, we calculate the total energy  per unit frequency radiated  by the charge on its path from $-\xi$ to $+ \xi$. Also we consider the ratio of the radiated energy per unit frequency between the $\theta_+$ and the $\theta_-$ cones, as well as between them  and the case of non-chiral materials $\theta_{\rm clas}$. We recall  the calculation in the conventional  Cherenkov case (non-chiral material), where 
we have 
\begin{equation}
\begin{split}
 \frac{d^2E_{\rm{clas}}}{d\Omega d\omega} & = \frac{n\omega^2q^2}{\pi^2c^3}\left( 1-\frac{c^2}{n^2v^2}\right) \frac{\sin^2[\xi \,\tilde{\Xi} ]}{\tilde{\Xi}^2}, \quad \tilde{\Xi}(\theta) = \frac{\omega}{v}\left(1-\frac{nv}{c}\cos\theta\right),
 \label{SPECDCONV} 
\end{split}
\end{equation}
given by  the limit $\alpha=0 $ in Eq.(\ref{SPECDIST1})
. Axial symmetry, which is also present in our chiral case, yields 
\begin{equation}
\frac{dE_{\rm{conv}}}{d\omega} = \frac{2n\omega^2 q^2}{\pi c^3}\left( 1-\frac{c^2}{n^2v^2}\right)\int_0^\pi d\theta \sin\theta \frac{\sin^2[\xi \,\tilde{\Xi} ]}{\tilde{\Xi}^2}.
\label{EDISTCONV}
\end{equation}
The trick to perform the remaining angular integration in the limit $\xi\rightarrow \infty$ is to use Eq. (\ref{ANGLECOND1}),  obtaining
\beq
\frac{\sin^2[\xi \,\tilde{\Xi} ]}{\tilde{\Xi}^2}=\pi \delta(\tilde{\Xi})\frac{\sin[\xi \,\tilde{\Xi} ]}{\tilde{\Xi}}=\pi \xi \delta(\tilde{\Xi}),
\label{TRICK}
\eeq
 leaving a final integration  over $\delta(\tilde{\Xi})$. The result is 
\begin{equation}
\frac{dE_{\rm{conv}}}{d\omega}= \frac{q^2 \omega L}{c^2} \left( 1-\frac{c^2}{n^2v^2}\right),
\label{DECLASSDW}
\end{equation}
where we have denoted by $L = 2\xi$ the total  distance traveled by the charged particle in the medium. Let us emphasize that in general $n=n(\omega)$, which we do not consider here. From Eq. (\ref{SPECDIST1}) we have
\begin{equation}
\frac{d^2 E}{d\Omega d\omega} =  \sum_{\eta= \pm 1}\frac{d^2E_{\eta}}{d\Omega d\omega}, \qquad  \frac{d^2E_{\eta}}{d\Omega d\omega} = \frac{n\omega^2q^2}{4\pi^2c^3} {\cal \tilde{K}}_\eta (\omega,\theta) \frac{\sin^2[\xi \tilde{\Xi}_\eta ]}{\tilde{\Xi}^2_\eta}, \qquad 
{\cal \tilde{K}}_\eta (\omega,\theta) = \frac{\mathcal{\tilde{T}}_{1,\eta}(\omega,\theta)}{\tilde{g}^2_\eta(\theta)},
\label{ANGDISCHIRAL}
\end{equation}
in the chiral case.
The angular integration is similar to the non-chiral case: the functions ${\cal \tilde{K}}_\eta(\omega, \theta)$ are evaluated at the respective Cherenkov angles $\theta_\eta$, while the integration over $\delta (\tilde{\Xi}_\eta)$ is a bit more involved. The result is 
\begin{equation}
\frac{dE_\eta}{d\omega}  = 
\frac{\omega q^2 L}{4 c^2} \left[ \frac{\sin\theta \, \cal{\tilde{K}}_\eta (\omega,\theta)}
{\Big|{\sin\theta \, \tilde{C}_\eta(\theta) - \cos\theta \, \frac{\partial \tilde{C}_\eta (\theta)}{\partial\theta}}\Big|}\right]_{\theta=\theta_\eta}.
\label{DEDOMEGAETA}
\end{equation}

\begin{figure}[h!]
\centering
\includegraphics[width=0.3 \textwidth]{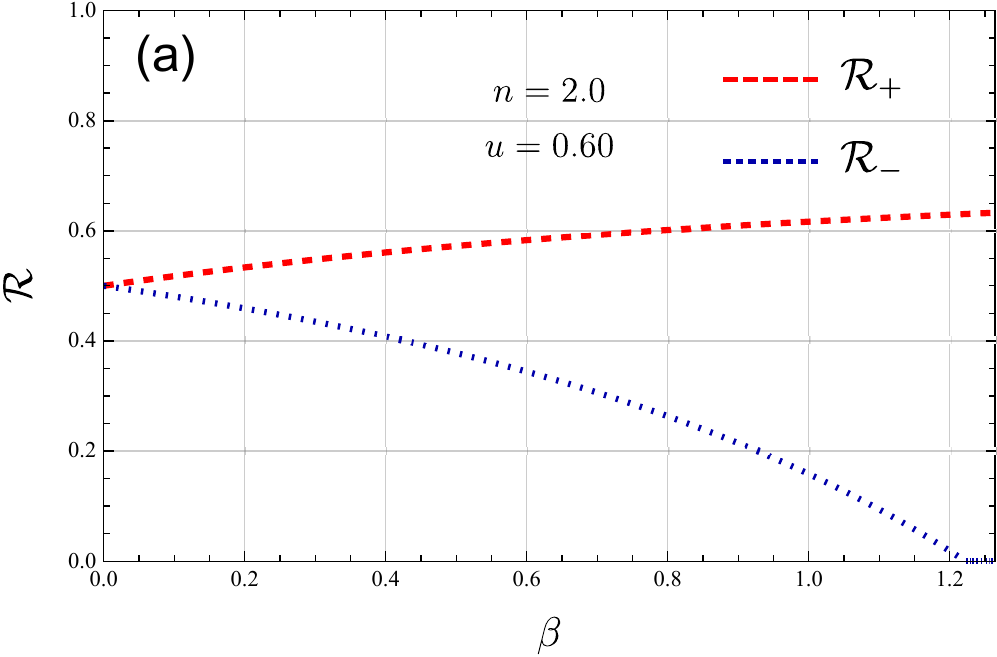}
\hspace{.3cm}
\includegraphics[width=0.3 \textwidth]{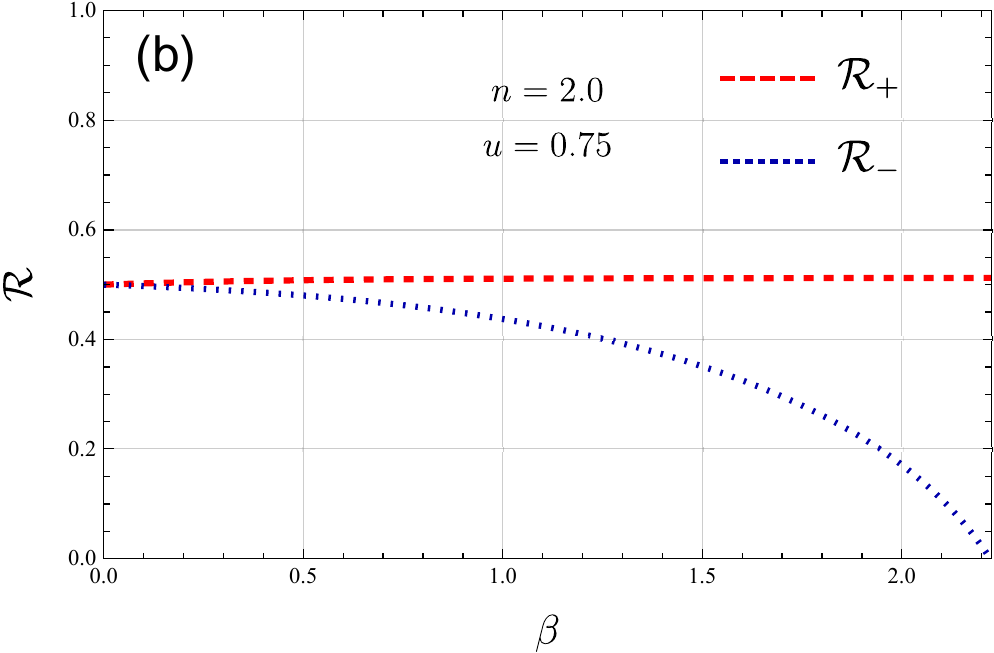}
\hspace{.3cm}
\includegraphics[width=0.3 \textwidth]{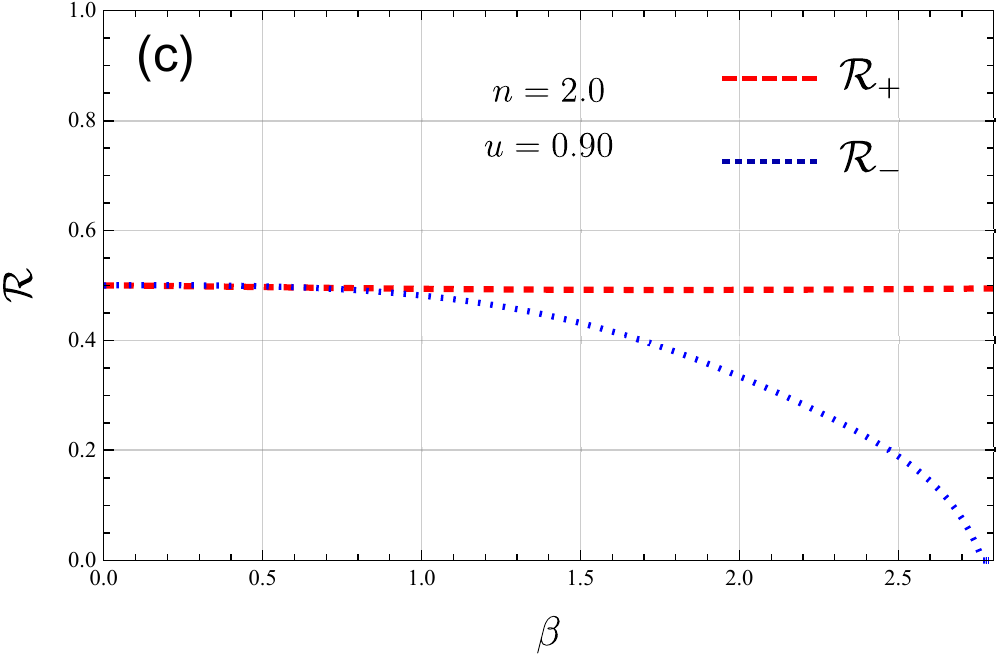}
\caption{Ratio between the total energy per unit frequency, as a function of $\beta$,  radiated by each cone ($\theta_+,\theta_-$ ) in the chiral Cherenkov case, with respect to the conventional case, for $n=2.0$. Panel (a): $u=0.6$, panel (b) : $u=0.75$ and panel (c) : $u= 0.9$. The dashed (red) line corresponds to ${\cal R}_+$ and the dotted (blue) line is for ${\cal R}_-$.}
\label{RATIOLOBEST}
\end{figure}

Now, we compare the contribution to the total energy radiated per unit frequency from each distribution in Eq. (\ref{DEDOMEGAETA}), as a function of the chiral parameter $\beta$. We define the ratios 
\begin{equation}
\mathcal{R}_\eta = \frac{dE_\eta/d\omega}{dE_{\text{class}}/d \omega} = \frac{1}{4}\frac{1}{1-\frac{c^2}{v^2n^2}} 
\left[ \frac{\sin\theta \,  \cal{\tilde{K}}_\eta (\omega,\theta)}
{\Big|{\sin\theta \, \tilde{C}_\eta(\theta) - \cos\theta \, \frac{\partial \tilde{C
}_\eta (\theta)}{\partial\theta}}\Big|}\right]_{\theta=\theta_\eta},
\label{RATIOS}
\end{equation}
which account for  the fraction of the radiated energy per unit frequency that is produced by each cone in the chiral case, with respect to the  energy per unit frequency emitted in the conventional  case. 

Figures \ref{RATIOLOBEST}(a-c)  show that when $\alpha_0$ is close to zero the contribution of the radiation for each $\eta$ in  Eq.\ (\ref{DEDOMEGAETA}) is approximately $0.5$.  As  $\beta$ takes larger values we see that the contribution of the radiation produced by  the inner cone, $\theta_-$, gets smaller and smaller until it disappears completely. The value of $\beta$ at which this occurs is given by the vanishing of   $\theta_-$  in Figs. \ref{FIG25} (b-d). 
In the limit when $\mathcal{R}_-$ vanishes, we observe that the contribution of the radiation from the outer cone  is just a fraction of the radiation in the conventional case.

\begin{figure}[h!]
\centering
\includegraphics[width=0.3 \textwidth]{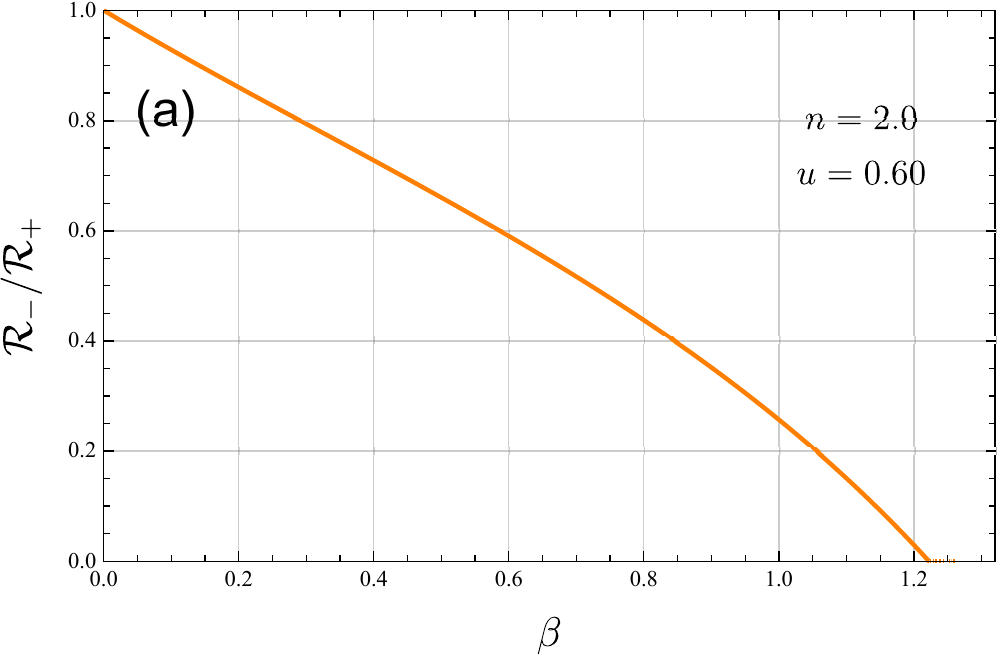}
\hspace{.3cm}
\includegraphics[width=0.3 \textwidth]{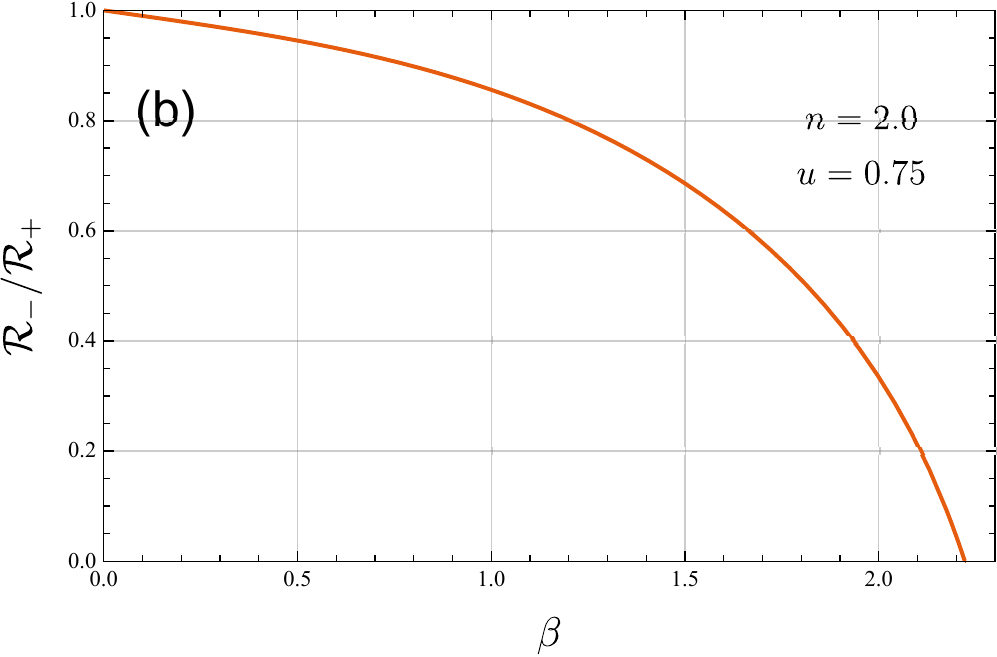}
\hspace{.3cm}
\includegraphics[width=0.3 \textwidth]{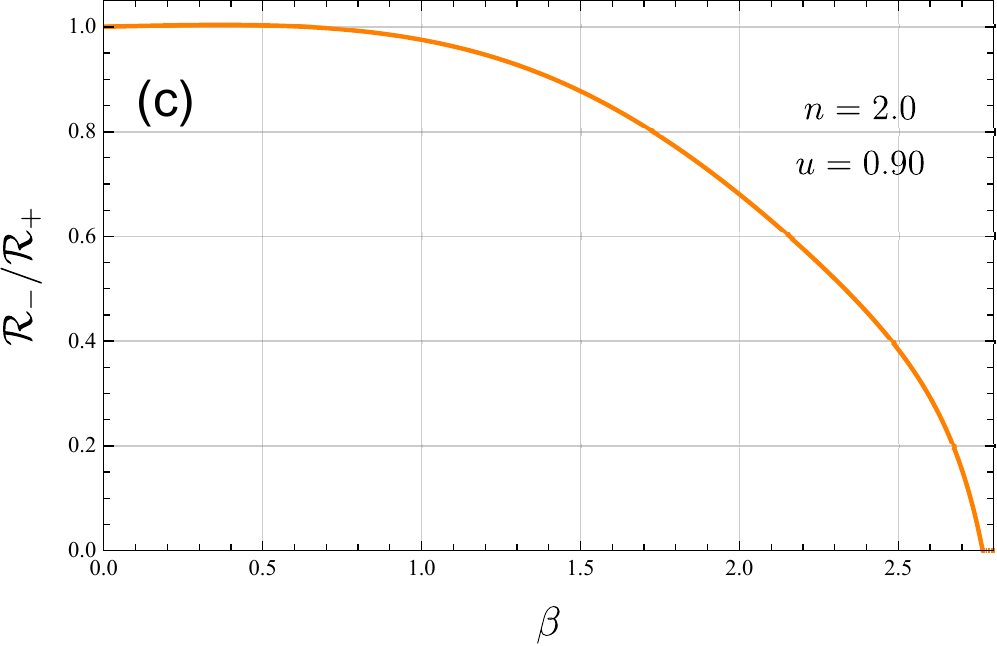}
\caption{Ratio between the total energy radiated  by the inner Cherenkov cone with respect to the outer one,  for $n=2.0$. {Panel (a) : $u=0.6$, panel (b) : $u=0,75$ and panel (c) : $u=0.9$.}}

\label{RATIOLOBEST1}
\end{figure}

To complete the analysis, we consider the ratio
\begin{equation}
\mathcal{R}_-/\mathcal{R}_+ = \frac{dE_-/d\omega}{dE_+/d \omega} = \frac{\sin\theta_- \, \cal{\tilde{K}}_- (\omega,\theta_-)}{\sin\theta_+ \,\cal{\tilde{K}}_+ (\omega,\theta_+)}   \frac{\Big|{\sin\theta\, \tilde{C}_+ - \cos\theta \, \frac{\partial \tilde{C}_+}{\partial\theta}}\Big|_{\theta=\theta_+}}{  \Big|{\sin\theta\,  \tilde{C}_- - \cos\theta \, \frac{\partial \tilde{C}_-}{\partial\theta}}\Big|_{\theta=\theta_-}},
\label{RRATIO}
\end{equation}
which is a measure of how larger the contribution to the radiation from the inner cone is with respect to the outer one. Figures \ref{RATIOLOBEST1}(a-c) show that the output  from the inner   cone is always smaller  than that from the outer cone. This ratio decreases as $\beta$ takes larger values, until it  gets to zero when  the inner cone vanishes.

\section{Summary and conclusions}

\label{SUM}

In the far field approximation, we study the electromagnetic response of a non magnetic chiral media to arbitrary   time-dependent sources in the framework of the Carroll, Field and Jackiw (CFJ) electrodynamics, which includes the additional parameters $b_0$ and $ \mbf{b}$  giving rise to the magnetoelectric effect. Examples of materials described by this effective electrodynamics are the Weyl semimetals, where 
these parameters describe the separation of the Weyl points in energy ($b_0$) and momentum ($\mbf{b}$), in the reciprocal space. It is interesting to remark that this model also arises in the CPT-odd sector of the Standard Model Extension designed to probe Lorentz invariance violations in fundamental interactions. We restrict ourselves to the case $b_0=0$ and choose the $z$ 
axis of our coordinate system in the direction of $\mbf{b}$. Most of our calculations are performed in vacuum ($n=1$) and the generalization to arbitrary ponderable media  required to deal with  a real material is performed following the substitutions indicated in the Appendix \ref{APPB}.

As discussed in section \ref{CHED}, the linear dependence $\theta(x)=b_\mu x^\mu$ on the coordinates in the CFJ electrodynamics preserves the invariance under translations of the  Lagrangian density up to a total derivative. The system has axial symmetry with respect to the 
vector $\mbf{b}$. In Eq. (\ref{FGCOMPL1}) we obtained the Green's function (GF) of the system,  $G^{\mu\nu}(x, x')$, in terms of a differential matrix operator acting on the scalar function $f(\mbf{x}, \mbf{x'},\omega)$, given  in Eq. (\ref{FCOMPL}). This function exhibits two contributions accounting for the birefringence  of the medium, which are inherited by the remaining observables.    The next step was to calculate  the general 
expression for the GF in the radiation zone. To this end we considered the stationary phase approximation in the far zone ($R \rightarrow \infty$), where the function $f(\mbf{x}, \mbf{x'},\omega)$ is highly oscillating according to 
$\exp(i R \, h_\eta(k_\perp))$ in Eqs. (\ref{OSC_PHAS}) and  (\ref{PHASE}). Still, the resulting  stationary phase equation (\ref{SPA}) is quartic in the momentum, which called for a further approximation to keep an analytic calculation. We studied the numerical solution of Eq. (\ref{SPA}) and concluded that, within 
specific ranges of the observation angle $\theta$, the choice in Eq.\ (\ref{APPK}), corresponding to the conventional case with $\mbf{b}=0$, is adequate. A qualitative measure of how good this approximation is for increasing values of $\beta=c |\mbf{b}|/\omega $ is shown in Fig. \ref{FIG01}. Yet at the end of the calculation of a Cherenkov angle one should identify its position in the corresponding graph in order to asses it validity. Next, for an arbitrary source  we determine the electromagnetic potentials in the radiation zone and give a general expression for the electromagnetic fields in Eqs. (\ref{FINB}) and (\ref{FINE}). We show that the triad ${\gv{\hat n}}, \mbf{E}, \mbf{B}$ is not orthogonal  and explain how the conventional case  is obtained. The general expression  for the 
spectral distribution of the radiation follows from the  Eqs. (\ref{EMPOT1}), (\ref{EFAAPP}), (\ref{IPMPM}) and (\ref{D2OMEGAFIN}). Some symmetry properties of the fields under the change 
$\omega \rightarrow -\omega$ are useful to show that this rather complicated expression is real, as expected. Also at each step in the calculation we able to recover the well known conventional results by setting $b=0$.

Next we turn to the particular case of Cherenkov radiation where  the source is a charge moving at constant velocity parallel to $\mbf{b}$, thus ignoring recoil effects. The electromagnetic potentials and fields results by the direct substitution of the charge current in the previous general expressions. In analogy with the conventional case, substantial simplifications arise from the form of the outgoing radiation wave which now acquires  additional angular dependence which modify the conventional Cherenkov condition. Yet this condition also arises as a delta-like contribution in the spectral distribution of the radiation which fixes the allowed angles according to the Eq.\ (\ref{ANGLECOND2}), which is conveniently written as $H_\eta(\theta)=c/nv$ as in Eq.\ (\ref{HTHETA}). Then, for given values of $b, n $ and $u=v/c$, the Cherenkov angles are determined by the intersection of 
$H_\eta$ with the straight line $1/nu$, as shown in Figs \ref{FIG1} and \ref{FIG2}, where we also plot the conventional case $H_{{\rm clas}}(\theta)$ for comparison. The functions $H(\theta)$ are decreasing in the angle. Figure \ref{FIG1} is for $n=1$ and depicts  what is normally called vacuum Cherenkov radiation that is forbidden in the conventional case. Here only the angle $\theta_+$ is present. Figure \ref{FIG2} is for $n>1$ and shows  zero, one or two Cherenkov angles. The limits of such regions is determined by the maximum  values
 $H(\theta=0)$ in the following way: (i) no Cherenkov angle occurs when $1/nu > H_+(0)$, (ii) when $1/nu <  H_+(0) $ the angle $\theta_+$  is always present and starts to be accompanied by the conventional angle when $1/nu <1$. The angle $\theta_-$ occurs only in the interval $ \alpha(n)(1-\alpha^2(n) < 1/nu <H_-(0)$. To avoid confusion, let us observe that the horizontal lines labeled by $u$ in those figures correspond to the value $1/nu$ in the ordinate.  The relation $\theta_- < \theta_{\rm clas}< \theta_+$ always hold and $\theta_{\pm}$ merge into $\theta_{\rm clas}$ as $|\mbf{b}| \rightarrow 0$.

We also study the dependence of the  angles $\theta_\pm$ as a function of the chiral parameter $\beta $ for fixed $n$ and $u$. As shown in Fig. \ref{FIG25}, we find two relevant cases: (1) for $nu<1$ we only have $\theta_+$ which starts at the minimum $\beta$ determined by the Eq.\ (\ref{NUMENUNO}), (2) for $nu>1$ both $\theta_\pm$ start at $\beta =0$. While $\theta_+$ is an increasing function, $\theta_-$ is a decreasing one which reaches zero at a value given by the Eq.\ (\ref{NUMAYUNO}).

Angular plots of the spectral distribution are presented in the Fig. \ref{FIG3}, showing the separation of the cones as ${\tilde \beta}$ increases. Also a qualitative measure of the energy output of each cone in given by the length of the radial lines, 
showing that $\theta_-$  radiates less than $\theta_+$. A better comparison is obtained by calculating the ratios  of the energy per unit frequency radiated in the $\theta_{\pm}$ 
 cones with respect to the conventional case (Fig. {\ref{RATIOLOBEST})}.  We find that the contribution of the outer cone ($\theta_+$) gets bigger, while  that of the  inner cone ($\theta_-$) gets smaller as $\beta = cb/\omega$ takes larger values, until it reaches the value zero. This behavior is also evident in Fig.\ \ref{RATIOLOBEST1}, which depicts the ratio of the energy radiated per unit frequency between the  $\theta_-$ and the $\theta_+$ cones.
 
We successfully compare our result for the vacuum Cherenkov angle $\theta_+$  with that obtained in Eq.\ (40) of Ref.\ 
\cite{TUCHIN3} in the high energy approximation. This requires
$\theta_+<<1$, $\beta = bc/\omega <<1 $, together with a highly relativistic velocity for the charge, which we take as $v = c$.  Under these conditions, our Eq.\ (\ref{ANGLECOND2}) reduces to 
\begin{equation}
\cos\theta_+\left[1+ \frac{1}{2}\beta \cos\theta_+ \right]= 1,
\end{equation}
whose solution yields $ \cos\theta_+ = 1 -{\beta}/{2}$, providing $\theta_+ = \sqrt{\beta}$ in the small angle approximation. This result is precisely that of Ref. \cite{TUCHIN3}, restricted to our case when $\mbf{v}$ is parallel to  $\mbf{b}$ and $\lambda=+$.

We expect that our general description of radiation in chiral matter provides the basis
for the study of further processes such as Cherenkov radiation with a 
charge velocity $\mbf{v}$ in arbitrary direction with respect to $\mbf{b}$, 
 charged particle energy losses and synchrotron 
radiation, for example. Regarding the case of Cherenkov radiation and without entering in any 
experimental detail, which is  far beyond our theoretical understanding, we envisage that the 
two Cherenkov angles predicted in this work
could be measured by sending the charge  through a slab of material and using either a differential Cherenkov counter or a ring imaging  Cherenkov
counter \cite{ratcliff}. It is interesting  to observe that the measure of $\theta_{+}$ would provide  an optical experimental determination of the parameter 
$|\mbf{b}|$ of  the chiral medium, according to the expression
\begin{eqnarray}
|\mbf{b}|=n^2 \left(\frac{\omega}{c}\right) \frac{1}{\cos^5\theta_+} \left(\frac{c}{nv}-\cos\theta_+ + 2 \cos^3 \theta_+ \right)\left(\frac{c}{nv}-\cos\theta_+ \right).
\end{eqnarray}
When both $\theta_{\pm}$ are present, $|\mbf{b}|$ can be determined from the measurement of  either one, and consistency yields  a relationship between $\theta_+$ and $\theta_-$, which is not very illuminating to be presented explicitly. 
 In our restricted formulation 
($\mathbf{v}$ parallel to $\mathbf{b}$) at least one should 
know the direction of $\mbf{b}$ in the sample to correctly send 
the incident charge parallel to $\mbf{b}$. 
Due to the far reaching applications of Cherenkov radiation and since chiral matter 
constitutes a genuine new state of matter recently discovered, it
could prove valuable to study its practical applications as a Cherenkov radiator.

\acknowledgments
{E.B.-A.} and L.F.U. acknowledge support from the project CONACyT CF/2019/428214. M.M.F. is supported by FAPEMA/ Universal/01187/18, FAPEMA/POS-GRAD 02575/21, CNPq/Produtividade 311220/2019-3, CNPq/Universal/ 422527/2021-1 and CAPES/Finance Code 001.

\appendix
\section{The calculation of the GF in the radiation zone}
\label{APPA}

In this Appendix  we summarize the calculation  of the GF  (\ref{FGCOMPL1}) in the far-field approximation, using the stationary phase method. We start from the splitting of the GF introduced in Eqs. (\ref{GF0}), (\ref{GFB}) and (\ref{GFB2}).

\subsection{$G^{ \mu \nu }_{0 }(\mbf{x},\mbf{x'}; \omega)$ in the far-field approximation}
Using $k^2= k_0^2 -k_\perp^2 +\partial_Z^2$ we further split $G^{ \mu \nu }_{0}(\mbf{x},\mbf{x'}; \omega) $ into
\begin{equation}
\begin{split}
G^{ \mu \nu }_{0}(\mbf{x},\mbf{x'}; \omega)  &  = - \eta^{\mu\nu} k_0^2 \frac{i}{16\pi b}\int^\infty_{-\infty} \frac{dk_\perp k_\perp}{k_\parallel} H_0^{(1)}(k_\perp R_\perp)
\left[ \frac{e^{i \sqrt{k_\parallel^2 +bk_\parallel } Z }}{\sqrt{k_\parallel^2 +bk_\parallel }} -\frac{e^{i\sqrt{k_\parallel^2 -bk_\parallel }Z}}{ \sqrt{k_\parallel^2 -bk_\parallel }}\right]   \\
& +\eta^{\mu\nu} \frac{i}{16\pi b}\int^\infty_{-\infty} \frac{dk_\perp k_\perp k^2_\perp}{k_\parallel} H_0^{(1)}(k_\perp R_\perp)
\left[ \frac{e^{i \sqrt{k_\parallel^2 +bk_\parallel } Z }}{\sqrt{k_\parallel^2 +bk_\parallel }} -\frac{e^{i\sqrt{k_\parallel^2 -bk_\parallel }Z}}{ \sqrt{k_\parallel^2 -bk_\parallel }}\right]  \\
& - \eta^{\mu\nu} \partial^2_Z \frac{i}{16\pi b}\int^\infty_{-\infty} \frac{dk_\perp k_\perp}{k_\parallel} H_0^{(1)}(k_\perp R_\perp)
\left[ \frac{e^{i \sqrt{k_\parallel^2 +bk_\parallel } Z }}{\sqrt{k_\parallel^2 +bk_\parallel }} -\frac{e^{i\sqrt{k_\parallel^2 -bk_\parallel }Z}}{ \sqrt{k_\parallel^2 -bk_\parallel }}\right].
\label{GFCOMPLETA1}\end{split}
\end{equation}
These three terms can be integrated using the stationary phase method in the same way as we have computed the function
\beq
f(\mbf{x},\mbf{x'};\omega)=\frac{i}{16\pi b}\int^\infty_{-\infty} \frac{dk_\perp k_\perp}{k_\parallel} H_0^{(1)}(k_\perp R_\perp)
\left[ \frac{e^{i \sqrt{k_\parallel^2 +bk_\parallel } Z }}{\sqrt{k_\parallel^2 +bk_\parallel }} -\frac{e^{i\sqrt{k_\parallel^2 -bk_\parallel }Z}}{ \sqrt{k_\parallel^2 -bk_\parallel }}\right],
\label{FAPP}
\eeq
in the far-field approximation, previously obtaining 
\begin{equation}
\frk{f}(\mbf{x},\mbf{x'};\omega)  = \frac{1}{8\pi r}  \frac{1}{k_0 b \cos\theta} \left[  \frac{e^{ik_0  C_+(\theta) (r-\hat{n}\cdot \mbf{x'}) }}{ g_+(\theta) } - \frac{e^{i k_0  C_-(\theta)(r-\hat{n}\cdot \mbf{x'}) }}{g_-(\theta) } \right], 
\label{FSPAAPP}
\end{equation}
in Eq. (\ref{FFIN}), with $k_0=\omega/c$.
Let us recall that  we are denoting by $\frk{f}(\mbf{x},\mbf{x'};\omega)$ the function ${f}(\mbf{x},\mbf{x'};\omega)$ evaluated in SPA and that according to our additional approximation in the SPA we set 
\beq
(k_\parallel)_s= k_0 \cos \theta, \qquad (k_\perp)_s=k_0 \sin \theta,
\label{RECALLSPP}
\eeq
in the final result.

The first integral in (\ref{GFCOMPLETA1}) is proportional to $f$ yielding the contribution 
\beq
-\eta^{\mu\nu}k^2_0 \, \frk{f}(\mbf{x},\mbf{x'};\omega),
\label{FTG00}
\eeq
in the SPA. The second one has an extra factor of $ k_\perp^2$ in the integrand, which  is evaluated  in the stationary phase point (\ref{RECALLSPP}), giving the contribution
\begin{equation}
\eta^{\mu\nu} k_0^2 \sin^2\theta \, \frk{f}(\mbf{x},\mbf{x'}; \omega).
\label{STG00}
\end{equation}
The third term in (\ref{GFCOMPLETA1}) requires the second derivative with respect Z, which is
\begin{equation}
\partial^2_Z \left[ \frac{e^{i \sqrt{k_\parallel^2 +bk_\parallel } Z }}{\sqrt{k_\parallel^2 +bk_\parallel }} -\frac{e^{i\sqrt{k_\parallel^2 -bk_\parallel }Z}}{ \sqrt{k_\parallel^2 -bk_\parallel }}\right] = - \left[ \sqrt{k_\parallel^2 +bk_\parallel }e^{i \sqrt{k_\parallel^2 +bk_\parallel } Z } -\sqrt{k_\parallel^2 -bk_\parallel } e^{i \sqrt{k_\parallel^2 -bk_\parallel }Z }  \right],
\label{DERZ2}
\end{equation}
and  can be rewritten as
\begin{equation}
\begin{split}
\eta^{\mu\nu}\frac{i}{16\pi b} & \int^\infty_{-\infty} \frac{dk_\perp k_\perp}{k_\parallel} H_0^{(1)}(k_\perp R_\perp)
\left[ (k_\parallel^2 +bk_\parallel) \frac{e^{i \sqrt{k_\parallel^2 +bk_\parallel } Z }}{\sqrt{k_\parallel^2 +bk_\parallel }}-(k_\parallel^2 -bk_\parallel) \frac{e^{i\sqrt{k_\parallel^2 -bk_\parallel }Z}}{ \sqrt{k_\parallel^2 -bk_\parallel }}  \right], \\
& = \eta^{\mu\nu} k_0^2 \cos^2\theta \Big[(1+\alpha\sec\theta)  \frk{f}_+(\mbf{x},\mbf{x'}; \omega)-(1-\alpha\sec\theta)  \frk{f}_-(\mbf{x},\mbf{x'}; \omega)\Big],
\end{split}
\label{TTG001}
\end{equation}
after evaluating the additional factors in the SPA. Substituting (\ref{STG00}) and (\ref{TTG001}) into (\ref{GFCOMPLETA1}), we finally obtain 
\ba
 && G^{\mu \nu }_{0}(\mbf{x},\mbf{x'}; \omega) =  \eta^{\mu \nu} \frac{1}{8\pi r}  \left [ \frac{  e^{ik_0  C_+(\theta)(r-\hat{n}\cdot \mbf{x'}) }}{g_+(\theta) } + \frac{  e^{i k_0  C_-(\theta)(r-\hat{n}\cdot \mbf{x'}) }}{ g_-(\theta) } \right].
\label{GFCOMPLETA3}
\ea

\subsection{$G^{\mu \nu }_{\text{\tiny{b}}}(\mbf{x},\mbf{x'}; \omega)$ in the far-field approximation}

Our starting point  is the contribution linear in $b$ in the second Eq.(\ref{FGCOMPL1}), which we rewrite  
\begin{equation}
G^{\;  \;  \mu \nu }_{\text{\tiny{b}} \;  \; }(\mbf{x},\mbf{x'}; \omega) =  i b \epsilon^{\mu \nu \sigma 3} \int  \frac{d^3 \mbf{k} }{(2\pi)^3} \frac{k_{\sigma}}{k^4 +\tilde{b}^2 k^2 - (\tilde{b} \cdot k )^2}e^{i \mbf{k}\cdot (\mbf{x}-\mbf{x'} )}.
\label{GFBAPP}
\end{equation}
Here $\sigma = 0,1,2$ and $k_\sigma = (k_0,-\mbf{k})$. It is convenient to define the following integral
\begin{equation}
I_\alpha(\mbf{x},\mbf{x'}; \omega)=   \int  \frac{d^3\mbf{k}}{(2\pi)^3} \frac{k_{\alpha}}{k^4 +\tilde{b}^2 k^2 - (\tilde{b} \cdot k )^2}e^{i \mbf{k}\cdot (\mbf{x}-\mbf{x'} )},
\label{IALPHA}
\end{equation}
such that 
\beq
G_b^{\mu\nu}=ib \epsilon^{\mu\nu\sigma 3} I_\sigma.
\label{GFBAPP1}
\eeq
 For $\sigma = 0$ we have straightforward result
\begin{equation}
I_0 (\mbf{x},\mbf{x'}; \omega)=k_0 \, \frk{f}(\mbf{x},\mbf{x'};\omega). 
\label{I0}
\end{equation}
Also, the contribution from $I_3 (\mbf{x},\mbf{x'}; \omega)$ yields zero in the GF  due to the Levi-Civita tensor.

Next we consider $\sigma= i=1,2$, and we perform the integration of  $k_z$ in the complex plane,  as we did previously in the calculation yielding Eq. (\ref{FCOMPL}). We are left with
\ba 
&&I^i (\mbf{x},\mbf{x'}; \omega)=  \ \frac{1}{(2\pi)^3}\int^\infty_0 dk_\perp k_\perp \int^{2\pi}_0 d\phi\,  k^{i} e^{ik_\perp R_\perp \cos\phi}\int^{\infty}_{-\infty}dk_z \frac{ e^{ik_z Z}}{ (k_0^2-\mbf{k}^2) ^2 - b^2 (k_0^2-\mbf{k}^2) - b^2 k_z^2}, \nonumber \\ 
&& \hspace{2.2cm}  = \frac{i}{16\pi^2 b}\int^\infty_0 \frac{dk_\perp k^2_\perp}{k_\parallel} \int^{2\pi}_0 d\phi \begin{bmatrix}
           \cos\phi \\
           \sin\phi
         \end{bmatrix}e^{ik_\perp R_\perp \cos\phi} \left[ \frac{e^{i \sqrt{k_\parallel^2 +bk_\parallel } Z }}{\sqrt{k_\parallel^2 +bk_\parallel }} -\frac{e^{i\sqrt{k_\parallel^2 -bk_\parallel }Z}}{ \sqrt{k_\parallel^2 -bk_\parallel }}\right],\nonumber \\
\label{II}
\ea
where we  recall  that $\phi$ is the angle between
${\mbf R}_\perp$ and ${\mbf k}_\perp$. In the second line of (\ref{II}) we have chosen a coordinate system $\cal{S}$ with the $x$-axis in the direction of ${\mbf R}_\perp$, such that $k^1 = k_\perp \cos\phi$ and $k^2= k_\perp \sin\phi$. 
The angular integrals are  
\beq
\int^{2\pi}_0 d\phi \cos\phi e^{ik_\perp R_\perp \cos\phi} = 2\pi i J_1(k_\perp R_\perp), \qquad \int^{2\pi}_0 d\phi \sin\phi e^{ik_\perp R_\perp \cos\phi} =0.
\label{ANGINT}
\eeq
Using the recurrence
\beq
J_1(k_\perp R_\perp) =-\frac{1}{k_\perp} \frac{\partial}{\partial R_\perp}J_0(k_\perp R_\perp),
\label{REC1}
\end{equation}
the integral (\ref{II})  can be written as
\begin{equation}
\begin{split}
\mbf{I}_\mathcal{S}(\mbf{x},\mbf{x'}; \omega) & = -\frac{1}{8\pi^2 b}\int^\infty_0 \frac{dk_\perp k^2_\perp}{k_\parallel} \begin{bmatrix}
           J_1(k_\perp R_\perp) \\
           0
         \end{bmatrix} \left[ \frac{e^{i \sqrt{k_\parallel^2 +bk_\parallel } Z }}{\sqrt{k_\parallel^2 +bk_\parallel }} -\frac{e^{i\sqrt{k_\parallel^2 -bk_\parallel }Z}}{ \sqrt{k_\parallel^2 -bk_\parallel }}\right], \\
& = \frac{\partial}{\partial R_\perp} \frac{1}{16\pi b}\int^\infty_{-\infty} \frac{dk_\perp k_\perp}{k_\parallel} \begin{bmatrix}
           H_0(k_\perp R_\perp) \\
           0
         \end{bmatrix} \left[ \frac{e^{i \sqrt{k_\parallel^2 +bk_\parallel } Z }}{\sqrt{k_\parallel^2 +bk_\parallel }} -\frac{e^{i\sqrt{k_\parallel^2 -bk_\parallel }Z}}{ \sqrt{k_\parallel^2 -bk_\parallel }}\right], \\
& = \frac{1}{i}  \begin{bmatrix}
           1 \\
           0
         \end{bmatrix}\frac{\partial}{\partial R_\perp}f(\mbf{x},\mbf{x'};\omega).       
\end{split}
\label{IS}
\end{equation}
In  the chosen coordinate system  we have  $I^2=0$. Then the vector $\mbf{I}(\mbf{x},\mbf{x'}; \omega) =(I^1, 0)$ is parallel to $\mbf{R_\perp}$ and we can write the general result
\begin{equation}
\mbf{I}(\mbf{x},\mbf{x'}; \omega)  =  \frac{1}{i} \frac{(\mbf{x}-\mbf{x'})_{\perp} }{R_\perp}\frac{\partial}{\partial R_\perp}  f(\mbf{x},\mbf{x'};\omega).
\label{IVECEQ}
\end{equation}
From the recurrence relations of the Hankel functions we obtain
\beq
 \frac{\partial}{\partial R_\perp} H_0^{(1)} (k_\perp R_\perp)= k_\perp  \frac{\partial}{\partial (k_\perp R_\perp)}H_0^{(1)} (k_\perp R_\perp)= - k_\perp H_1^{(1)} (k_\perp R_\perp),
 \label{REC2}
\eeq
which yields
\begin{equation}
\frac{\partial}{\partial R_\perp}  f(\mbf{x},\mbf{x'};\omega) = -\frac{i}{16\pi b}\int^\infty_{-\infty} \frac{dk_\perp k^2_\perp}{k_\parallel} H_1^{(1)}(k_\perp R_\perp)
\left[ \frac{e^{i \sqrt{k_\parallel^2 +bk_\parallel } Z }}{\sqrt{k_\parallel^2 +bk_\parallel }} -\frac{e^{i\sqrt{k_\parallel^2 -bk_\parallel }Z}}{ \sqrt{k_\parallel^2 -bk_\parallel }}\right].
\label{DERFAPP}
\end{equation}
In the asymptotic limit $ k_\perp R_\perp \rightarrow \infty$, we can approximate \cite{abramowitz} 
\begin{equation}
H_1^{(1)}(k_\perp R_\perp)  = \sqrt{\frac{2}{\pi k_\perp R_\perp}}e^{ik_\perp R_\perp -i\frac{\pi}{4}} e^{-i\frac{\pi}{2}}= H_0^{(1)}(k_\perp R_\perp)  e^{-i\frac{\pi}{2}} = -i H_0^{(1)}(k_\perp R_\perp),
\label{ASYMP}
\end{equation}
\begin{equation}
\frac{\partial}{\partial R_\perp}  f(\mbf{x},\mbf{x'};\omega)  = i \frac{i}{16\pi b}\int^\infty_{-\infty} \frac{dk_\perp k^2_\perp}{k_\parallel} \, H_0^{(1)}(k_\perp R_\perp)\, 
\left[ \frac{e^{i \sqrt{k_\parallel^2 +bk_\parallel } Z }}{\sqrt{k_\parallel^2 +bk_\parallel }} -\frac{e^{i\sqrt{k_\parallel^2 -bk_\parallel }Z}}{ \sqrt{k_\parallel^2 -bk_\parallel }}\right], \\
\label{DERFAPP1}
\end{equation}
which again has the form of the function $f(\mbf{x},\mbf{x'};\omega)$
in Eq.\ (\ref{FAPP}), except for an additional factor $k_\perp$ in the integrand. In the SPA this implies
\beq
\frac{\partial}{\partial R_\perp}  f(\mbf{x},\mbf{x'};\omega)=ik_0 \sin \theta \, \frk{f}(\mbf{x},\mbf{x'};\omega), \quad \rightarrow \quad 
\mbf{I}(\mbf{x},\mbf{x'};\omega) = k_0  \hat{\mbf{n}}_\perp   \frk{f}(\mbf{x},\mbf{x'};\omega),
\label{FINALIVEC}
\eeq
where $\hat{\mbf{n}}_{\perp}= (\sin\theta \cos \phi, \sin\theta \sin\phi)$. This completes the calculation of $G_b^{\mu\nu}$ in Eq.\ (\ref{GFB}).

\subsection{$G^{\mu\nu  }_{\text{\tiny{b}}^2 }(\mbf{x},\mbf{x'}; \omega)$ in the far-field approximation}
This term is the easiest to calculate, because it is proportional to $  f(\mbf{x},\mbf{x'};\omega)$, with
\begin{equation}
G^{\mu\nu  }_{\text{\tiny{b}}^2 }(\mbf{x},\mbf{x'}; \omega)  = -\tilde{b}^\mu \tilde{b}^\nu f(\mbf{x},\mbf{x'};\omega).
\label{GFB2APP}
\end{equation}
The only non-zero contribution in SPA is 
\begin{equation}
G^{3 3  }_{\text{\tiny{b}}^2 }(\mbf{x},\mbf{x'}; \omega) = -b^2 \frk{f} (\mbf{x},\mbf{x'};\omega).
\label{GFB233}
\end{equation}

\section{Electromagnetic fields and the spectral distibution of the radiation for $n=1$}
\label{APPB}

First we write the Cartesian components of $A^\mu$ for an arbitrary current $J^\mu (\mbf{x'},  \omega)$ in terms of the Fourier transform ${\cal J}^\mu({\mbf k}_\eta, \omega)$, defined in Eq.\ (\ref{INT}), which we denote as ${\cal J}^\mu_\eta$. From Eq.\ (\ref{APART})  and in the notation of Eq.\ (\ref{EMPOT2}) we have
\ba
&& {\cal A}^{0}_\eta (\mbf{\hat{n}},\omega) = 
{\cal J}^{0}_\eta + i \eta \tan\theta\sin\phi \, {\cal J}^{1}_\eta- i\eta \tan \theta \cos\phi \, {\cal J}^{2}_\eta,\nonumber \\
&& {\cal A}^{1}_\eta (\mbf{\hat{n}},\omega) =  i \eta \tan\theta \sin\phi \,  {\cal J}^{0}_\eta + {\cal J}^{1}_\eta - i \eta \sec\theta \, {\cal J}^{2}_\eta, \nonumber \\
&& {\cal A}^{2}_\eta (\mbf{\hat{n}},\omega) =  - i \eta \tan\theta \cos \phi \,  {\cal J}^{0}_\eta + i \eta\sec\theta \,  {\cal J}^{1}_\eta + {\cal J}^{2}_\eta,\nonumber \\
&& {\cal A}^{3}_\eta (\mbf{x},\omega) = (1 + \eta \, \beta \sec\theta){\cal J}^{3}_\eta.
\label{CALAJ}
\ea
The transformation to spherical coordinates yields
\begin{eqnarray}
{\cal A}^r_\eta(\mbf{\hat{n}},\omega) & = & \sin\theta\cos\phi {\cal A}^1_\eta + \sin\theta \sin\phi {\cal A}^2_\eta + \cos\theta {\cal A}^3_\eta, \nonumber\\
{\cal A}^\theta_\eta(\mbf{\hat{n}},\omega) & = &\cos\theta\cos\phi  {\cal A}^1_\eta + \cos\theta\sin\theta  {\cal A}^2_\eta -\sin\theta  {\cal A}^3_\eta,  \\
{\cal A}^\phi_\eta(\mbf{\hat{n}},\omega) & = & -\sin\theta  {\cal A}^1_\eta +\cos\phi  {\cal A}^2_\eta. \label{ASPHAPP}
\end{eqnarray}
Using Eq.\ (\ref{GRADRAD}) for the expression of the gradient operator in the radiation zone in the conventional relations 
\beq
\mbf{E}(\mbf{x},\omega) = ik_0 \mbf{A}(\mbf{x},\omega) -\gv{\nabla} A^{0}(\mbf{x},\omega) , \qquad 
\mbf{B}(\mbf{x},\omega)  =  \gv{\nabla} \times \mbf{A}(\mbf{x},\omega), \label{EMFAPP}
\eeq
and splitting the electromagnetic fields as shown in Eqs. (\ref{FINB}) and (\ref{FINE}) we have
\ba
&&\mbf{E}_\eta(\mbf{x},\omega) =i k_0 \left( \mbf{A}_\eta(\mbf{x},\omega)- 
{\mbf N}_\eta A^0_\eta(\mbf{x},\omega) \right),
\label{EFINAPP}\\
&& \mbf{B}_\eta(\mbf{x},\omega)  
 = {\mbf N}_\eta \times \mbf{E}_\eta(\mbf{x},\omega), 
 \label{BFINAPP}
\ea
recalling the definition of ${\mbf N}_\eta$ in Eq.\ (\ref{DEFN}).
The explicit form of the  radiation electric field in spherical coordinates is 
\beq
\mbf{E}_\eta(\mbf{x},\omega) = i k_0 \left[ \left( A^r_\eta-C_\eta A^0_\eta \right) \hat{{\mbf r}} +\left( A^\theta_\eta-\frac{\partial C_\eta}{\partial \theta}A^0_\eta\right)\gv{\hat{\theta}} + A^\phi_\eta \,  \gv{\hat{\phi}} \right],
\label{ESPHERAPP}
\eeq 
Now we are in position to calculate the  spectral energy  distribution (SED) of the radiation: the  energy radiated per unit solid angle and per unit frequency. We have taken $b_0=0$ which yield the Poynting vector 
\begin{equation}
\mbf{S}(\mbf{x},t) = \frac{c}{4\pi}\mbf{E}(\mbf{x},t)\times \mbf{B}(\mbf{x},t).
\label{POYTINGAPP}
\end{equation}
Thus, the total radiated energy crossing the  area $\hat{{\mbf r}} \, dA$ is  
\begin{equation}
\begin{split}
\int_{-\infty}^\infty dt (\hat{\mbf{r}}\cdot\mbf{S}(\mbf{x},t))\, dA &= \frac{c}{4 \pi} \int_{-\infty}^\infty \frac{d\omega}{2 \pi} \, 
\hat{\mbf{r}}\cdot \left(  \mbf{E}^*(\mbf{x},\omega) \times \mbf{B}(\mbf{x},\omega)  \right) \,dA,   \\
&= \frac{c}{4 \pi^2} \int_0^\infty d \omega \, \hat{\mbf{r}}\cdot {\rm Re}
\Big[  \mbf{E}^*(\mbf{x},\omega) \times \mbf{B}(\mbf{x},\omega) \Big] r^2 d \Omega, \label{TOTRADE}
\end{split} 
\end{equation}
where $d A= r^2 d \Omega$ and we read the SED
\beq
\frac{d^2 E}{d \Omega d \omega}=
\frac{c}{4 \pi^2} r^2 \, \hat{\mbf{r}}\cdot  {\rm Re}
\Big[ \mbf{E}^*(\mbf{x},\omega) \times \mbf{B}(\mbf{x},\omega)  \Big]. 
\label{SPECTDISAPP}
\eeq

The last step in Eq.\ (\ref{TOTRADE}) is a consequence of the relation 
\beq
\mbf{E}^*(\mbf{x},-\omega) \times \mbf{B}(\mbf{x},-\omega)=
\mbf{E}(\mbf{x},\omega) \times \mbf{B}^*(\mbf{x}, \omega)
=(\mbf{E}^*(\mbf{x},\omega) \times \mbf{B}(\mbf{x},\omega))^*,
\label{SYMOMEGAAPP}
\eeq
in the integral   $-\infty < \omega < +\infty$, which follows  because the electromagnetic fields are real in coordinate space. Now we  calculate the term
\begin{equation}
\mbf{E}^*(\mbf{x},\omega) \times \mbf{B}(\mbf{x},\omega)  =\sum_{\eta,\eta'= \pm 1}
 \mbf{E}^*_{\eta'}(\mbf{x},\omega) \times \mbf{B}_{\eta}(\mbf{x},\omega).
 \label{ECROSSB}
\end{equation}
From Eq.\ (\ref{BFINAPP}) we have
\begin{equation}
\mbf{E}^*_{\eta'} \times \mbf{B}_{\eta} =\mbf{E}^*_{\eta'}(\mbf{x},\omega) \times \left[  {\mbf N}_{\eta}  \times  \mbf{E}_{\eta}(\mbf{x},\omega) \right] \\
 = (\mbf{E}^*_{\eta'} \cdot \mbf{E}_{\eta}){ \mbf N}_\eta-(\mbf{E}^*_{\eta'}\cdot {\mbf N}_\eta) \mbf{E}_\eta.
 \label{ECROSSBETA}
\end{equation}
Fom Eq.\ (\ref{ESPHERAPP}), the electric field can be rewritten  in   spherical components as
\ba
&&\mbf{E}_\eta (\mbf{x},\omega) = ik_0\Big[\mathcal{E}^r_\eta\hat{{\mbf r}}+\mathcal{E}^\theta_\eta  \gv{\hat{\theta}}+\mathcal{E}^\phi_\eta\gv{\hat{\phi}}\Big], \nonumber \\
&&  \mbf{E}^{*}_{\eta'} (\mbf{x},\omega) = -ik_0\Big[(\mathcal{E}^r_{\eta'})^* \hat{{\mbf r}}+(\mathcal{E}^\theta_{\eta'})^*  \gv{\hat{\theta}}+(\mathcal{E}^\phi_{\eta'})^*\gv{\hat{\phi}}\Big],
\label{EETASPH}
\ea
where
\begin{eqnarray}
\mathcal{E}_\eta^r & = &  A^r_\eta-C_\eta A^0_\eta ,\qquad
\mathcal{E}_\eta^\theta =   A^\theta_\eta-\frac{\partial C_\eta}{\partial \theta}A^0_\eta, \qquad 
\mathcal{E}_\eta^\phi = A^\phi_\eta. 
\label{EFAAPP}
\end{eqnarray}
With these we compute
\ba
&&\mbf{E}^*_{\eta'} \cdot {\mbf N}_\eta=\mbf{E}^*_{\eta'} \cdot \left( C_\eta \gv{\hat{r}}+\frac{\partial C_\eta}{\partial \theta} \gv{\hat{\theta}}\right)=- ik_0 \left[(\mathcal{E}^r_{\eta'})^*C_\eta+ (\mathcal{E}^\theta_{\eta'})^* \frac{\partial C_\eta}{\partial\theta}\right], \label{EDOTNAPP}\\
&& \mbf{E}^*_{\eta'} \cdot \mbf{E}_\eta= k_0^2 \Big[(\mathcal{E}^r_{\eta'})^*\mathcal{E}^r_\eta+(\mathcal{E}^\theta_{\eta'})^*\mathcal{E}^\theta_\eta+(\mathcal{E}^\phi_{\eta'})^*\mathcal{E}^\phi_\eta\Big],
\label{EDOTEAPP}
\ea
yielding 
\begin{equation}
\begin{split}
\mbf{E}^*_{\eta'} \times \mbf{B}_{\eta} & = k_0^2 \Big[(\mathcal{E}^r_{\eta'})^*\mathcal{E}^r_{\eta}+(\mathcal{E}^\theta_{\eta'})^*\mathcal{E}^\theta_{\eta}+(\mathcal{E}^\phi_{\eta'})^*\mathcal{E}^\phi_{\eta}\Big]\left(C_{\eta}\hat{{\mbf r}}+\frac{\partial C_{\eta}}{\partial \theta}\gv{\hat{\theta}}\right)
  \\
& -k_0^2 \left[(\mathcal{E}^r_{\eta'})^*C_{\eta}+ (\mathcal{E}^\theta_{\eta'})^* \frac{\partial C_{\eta}}{\partial\theta}\right] \Big[\mathcal{E}^r_{\eta} \hat{{\mbf r}}+\mathcal{E}^\theta_{\eta}\gv{\hat{\theta}}+\mathcal{E}^\phi_{\eta}\gv{\hat{\phi}}\Big].
\label{EDOTBETA}
\end{split}
\end{equation}
From the above equation we obtain 
\begin{equation}
{\hat{{\mbf r}}} \cdot (\mbf{E}^*_{\eta'} \times \mbf{B}_m)  =
 k_0^2 \left[\Big( (\mathcal{E}^\theta_{\eta'})^*\mathcal{E}^\theta_\eta+(\mathcal{E}^\phi_{\eta'})^*\mathcal{E}^\phi_\eta \Big) C_\eta - (\mathcal{E}^\theta_{\eta'})^* 
 \mathcal{E}^r_\eta\frac{\partial C_\eta}{\partial\theta}  \right] \equiv {\cal I}_{\eta' \eta}(\hat{{\mbf r}}, \omega)),
 \label{RDOTPOYETA}
\end{equation}
which produces the final result
\begin{equation}
\hat{{\mbf r}}\cdot (\mbf{E}^*(\mbf{x},\omega) \times \mbf{B}(\mbf{x},\omega))  =\sum_{\eta,\eta'= \pm 1}
 {\cal I}_{\eta' \eta}(\hat{{\mbf r}}, \omega)),
 \label{RDOTPOY}
\end{equation}
with the explicit expressions
\begin{eqnarray}
{\cal I}_{++}(\hat{{\mbf r}}, \omega)) & = &  k_0^2 \left[\Big( (\mathcal{E}^\theta_+)^*\mathcal{E}^\theta_++(\mathcal{E}^\phi_+)^*\mathcal{E}^\phi_+\Big) C_+ - (\mathcal{E}^\theta_+)^* \mathcal{E}^r_+\frac{\partial C_+}{\partial\theta}  \right], \nonumber \\
{\cal I}_{--}(\hat{{\mbf r}}, \omega)) & = & k_0^2 \left[\Big( (\mathcal{E}^\theta_-)^*\mathcal{E}^\theta_-+ (\mathcal{E}^\phi_-)^*\mathcal{E}^\phi_-\Big) C_- - (\mathcal{E}^\theta_-)^* \mathcal{E}^r_-\frac{\partial C_-}{\partial\theta}  \right], \nonumber \\
{\cal I}_{+-}(\hat{{\mbf r}}, \omega)) & = &  k_0^2 \left[\Big( (\mathcal{E}^\theta_+)^*\mathcal{E}^\theta_-+ (\mathcal{E}^\phi_+)^*\mathcal{E}^\phi_-\Big) C_- - (\mathcal{E}^\theta_+)^* \mathcal{E}^r_-\frac{\partial C_-}{\partial\theta}  \right],\nonumber\\
{\cal I}_{-+}(\hat{{\mbf r}}, \omega)) & = &  k_0^2 \left[\Big( (\mathcal{E}^\theta_-)^*\mathcal{E}^\theta_++ (\mathcal{E}^\phi_-)^*\mathcal{E}^\phi_+\Big) C_+ - (\mathcal{E}^\theta_-)^* \mathcal{E}^r_+\frac{\partial C_+}{\partial\theta}  \right].\label{IPMPM}
\end{eqnarray}
{Thus, by substituting  (\ref{RDOTPOY}) into (\ref{SPECTDISAPP}), it
 provides} the final expression
\begin{equation}
 \frac{d^2E}{d\Omega d\omega} = \frac{c}{4\pi^2} r^2 \, {\rm Re}\Big[\sum_{\eta,\eta'= \pm 1}
 {\cal I}_{\eta' \eta}(\gv{\hat{r}},\omega) \Big],
\label{D2OMEGAFIN}
\end{equation}
for the SED of the radiation in terms of the electromagnetic fields for arbitrary sources $J^\mu(\mbf{x}, t)$ in the case $n=1$.

Using the definitions  (\ref{EFAAPP}), Eqs. (\ref{SYMCG}) and (\ref{SYMAMU}) allow to obtain the symmetry relations
\beq
\mathcal{E}_\eta^r(-\omega) =(\mathcal{E}_{-\eta}^r(\omega) )^*, \qquad 
\mathcal{E}_\eta^\theta(-\omega) = (\mathcal{E}_{-\eta}^\theta(\omega))^*, \quad 
\mathcal{E}_\eta^\phi(-\omega)=(\mathcal{E}_{-\eta}^\phi(\omega))^*,
\label{SYMEAPP}
\eeq
which then yield
\begin{equation}
\mathcal{I}_{++}(\hat{{\mbf r}},-\omega)= (\mathcal{I}_{--}(\hat{{\mbf r}}, \omega))^*, \qquad {\cal I}_{+-}(\hat{{\mbf r}},-\omega)=({\cal I}_{-+}(\hat{{\mbf r}},\omega))^*.
\label{SYMIPMPM}
\end{equation}
To conclude  we summarize additional symmetry properties used along the text, resulting from  the transformation  $\omega \rightarrow - \omega$, which produces $\beta \rightarrow - \beta$, that in turn can be absorbed in the change $\eta \rightarrow -\eta, \, $  together with  complex conjugation in some cases
\ba
&& J^\mu(\mbf{x}, -\omega)=(J^\mu(\mbf{x}, \omega))^*, \qquad A^\mu_\eta(\mbf{x}, -\omega)=(A^\mu_{-\eta}(\mbf{x}, \omega))^*, \nonumber \\
&&\tilde{C}_\eta (-\omega, \theta) =\tilde{C}_{- \eta } (\omega, \theta), \qquad 
\tilde{g}_\eta (-\omega, \theta) =\tilde{g}_{- \eta} (\omega, \theta).
\label{SYMPROP}
\ea

\section{The spectral distribution in the Cherenkov radiation for $n=1$}
\label{APPC}

We present only the main steps in the calculation  of the SED for the Cherenkov radiation. We start from the sources in frequency space
\begin{eqnarray}
\rho (\mathbf{x}^{\prime},\omega) = \frac{q}{v} \delta(x^{\prime})
\delta(y^{\prime})e^{i \omega \frac{z^{\prime}}{v}}, \qquad {J}^3(%
\mathbf{x}^{\prime},\omega) = q \delta(x^{\prime}) \delta(y^{\prime})e^{i
\omega \frac{z^{\prime}}{v}}, \quad J^1=J^2=0, \label{SOURCESCHAP}
\end{eqnarray}
describing a charge $q$ moving in the third direction $z$  with constant speed $v$.
In order to have a well defined limiting process in our calculation we
follow Refs. \cite{Panofsky,PRDOJF} integrating the charge
trajectory in the interval $z\in (-\xi ,\xi )$ and taking the limit $\xi
\rightarrow \infty $ at the end of the calculation. 
In terms of the Fourier transform ${\cal J}^\mu({\mbf k}_\eta, \omega)$, defined in Eq.\ (\ref{INT}), we obtain the vector potential ${\cal A}_\mu$ from Eqs. (\ref{CALAJ}) and (\ref{ASPHAPP}) which yield the electric field (\ref{ESPHERAPP}), from where we read the components ${\cal E}_\eta^r,\, {\cal E}_\eta^\theta, \, {\cal E}_\eta^\phi $, according to Eq. (\ref{EFAAPP})
\begin{eqnarray}
\mathcal{E}^r_\eta & = &  \frac{q}{c r}   \frac{\sin[\xi \,\Xi_\eta ]}{\Xi_\eta}\left[ \cos\theta +\eta \beta -
\frac{c}{v}C_\eta(\theta)\right] \frac{e^{ik_0  r C_\eta (\theta)}}{ g_\eta(\theta) },    \\
\mathcal{E}^\theta_\eta & = & -\frac{q}{c r}   \frac{\sin[\xi 
\, \Xi_\eta ]}{\Xi_\eta}\left[ \sin\theta + \eta \beta \tan\theta + \frac{c}{v}\frac{\partial C_\eta (\theta)}{\partial\theta}\right] \frac{e^{ik_0  r C_\eta (\theta)}}{ g_\eta (\theta) }, \\
\mathcal{E}^\phi_\eta & = &  -\frac{q}{c r}   \frac{\sin[\xi \, \Xi_\eta ]}{\Xi_\eta}\left[ i \,\eta \frac{c}{v}\tan\theta \right] \frac{e^{ik_0 r C_\eta (\theta)}}{ g_\eta (\theta) }.
\end{eqnarray}
The expressions for  $C_\eta(\theta)$ and $g_\eta(\theta)$ are given in Eqs. (\ref{CETA}) and (\ref{GETA}), respectively, and we have
\begin{equation}
\Xi_\eta=\frac{\omega}{v}\Big(1-\frac{v}{c} C_\eta(\theta)\cos \theta   \Big).
\end{equation}
Finally, we substitute  these expressions for the components of the electric field in Eqs. (\ref{IPMPM}) to obtain 
 
\begin{eqnarray}
\mathcal{I}_{++}(\gv{\hat{r}},\omega) & = & \frac{k_0^2q^2}{c^2r^2}\frac{\sin^2[\xi \,\Xi_+ ]}{\Xi^2_+} \mathcal{T}_{1,+}(\omega,\theta) \frac{1}{g^2_+(\theta)},\nonumber \\
\mathcal{I}_{--}(\gv{\hat{r}},\omega)  & = & \frac{k_0^2q^2}{c^2r^2}\frac{\sin^2[\xi \,\Xi_- ]}{\Xi^2_-} \mathcal{T}_{1,-}(\omega,\theta)  \frac{1}{g^2_-}(\theta),\nonumber \\
\mathcal{I}_{+-}(\gv{\hat{r}},\omega)   & = & \frac{k_0^2q^2}{c^2r^2}\frac{\sin[\xi \, \Xi_+ ]}{\Xi_+} \frac{\sin[\xi \, \Xi_- ]}{\Xi_-}\mathcal{T}_{2,+}(\omega,\theta)  \frac{e^{ik_0 r(C_-(\theta)-C_+(\theta))}}{g_+(\theta)g_-(\theta)}, \nonumber\\
\mathcal{I}_{-+}(\gv{\hat{r}},\omega)   & = & \frac{k_0^2q^2}{c^2r^2}\frac{\sin[\xi \, \Xi_+ ]}{\Xi_+} \frac{\sin[\xi \,\Xi_- ]}{\Xi_-} \mathcal{T}_{2,-}(\omega,\theta)  \frac{e^{-ik_0 r(C_-(\theta)-C_+(\theta))}}{g_+(\theta)g_-(\theta)},\label{I4cherenkovrad}
\end{eqnarray}
The functions  ${\cal{T}}_{1,\eta}$ and  ${\cal { T}}_{2,\eta}$ are written in a compact by introducing the  the auxiliary quantities
\ba
&& {p}_\eta(\omega, \theta)= \sin\theta + \eta { \beta} \tan\theta
+ \frac{c}{v}\frac{\partial {C}_\eta}{\partial\theta} , \qquad {q}_\eta(\omega, \theta)= \cos\theta+ \eta {\beta}  -\frac{c}{v}{C}_\eta, 
\label{PQETAAP} 
\ea
and they read 
\ba
&& \mathcal{{T}}_{ 1,\eta }  =\left( {p}^2_\eta + \frac{c^2}{ v^2}\tan^2\theta \right){C}_\eta 
+ {p}_\eta {q}_{\eta}\frac{%
\partial {C}_\eta}{\partial\theta},\quad 
\mathcal{{T}}_{2,\eta}  =\left({p}_-{p}_+ - 
\frac{c^2}{ v^2}\tan^2\theta \right){C}_{-\eta} +{p}_\eta {q}_{-\eta} \frac{%
\partial {C}_{-\eta}}{\partial\theta}.\nonumber \\
\label{CALTETAAP} 
\ea
The symmetry properties of the remaining functions are 
\begin{eqnarray}
&& \tilde{\Xi}_\eta(-\omega,\theta) =-\tilde{\Xi}_{-\eta}(\omega,\theta), \qquad \tilde{p}_\eta(-\omega, \theta)=\tilde{p}_{-\eta} (\omega, \theta), \qquad \tilde{q}_\eta(-\omega, \theta)= 
\tilde{q}_{- \eta}(\omega, \theta), \nonumber \\
&& \mathcal{\tilde{T}}_{1, \eta}(-\omega,\theta)= \mathcal{\tilde{T}}_{ 1, -\eta}(\omega,\theta), \qquad 
\mathcal{\tilde{T}}_{2,\eta}(-\omega,\theta)= \mathcal{\tilde{T}}_{2,-\eta}(\omega,\theta).
\end{eqnarray}
   
The final expression for the SED is obtained by substituting the relations (\ref{I4cherenkovrad}) in Eq. (\ref{D2OMEGAFIN}).

\section{Introducing  ponderable media with parameters  $\epsilon$ and $\mu$}
\label{APPD}

In this Appendix we relate the physical Maxwell equations describing chiral matter for a non-dispersive, non-disipative medium with those used in the manuscript given in Eq.\  (\ref{MAX1}) and (\ref{MAX2}).
In  unrationalized Gaussian units the Maxwell equations for an  ideal chiral medium with $\epsilon=\mu=1$ are
\ba
&&\gv{\nabla} \cdot \mathbf{E}^{\prime }=\frac{4\pi}{c'}J^{^{\prime }0}-\mathbf{\ b}^{\prime
}\cdot \mathbf{B}^{\prime },\qquad \gv{\nabla} \times \mathbf{B}^{\prime }-\frac{1%
}{c^{\prime }}\frac{\partial \mathbf{E}^{\prime }}{\partial t}=\frac{4\pi}{%
c^{\prime }}\mathbf{J}^{\prime }+{\ b}_{0}^{\prime }\mathbf{B}^{\prime }+%
\mathbf{\ b}^{\prime }\times \mathbf{E}^{\prime}, \nonumber \\
&&
\gv{\nabla} \cdot \mathbf{B}^{\prime }=0,\qquad \gv{\nabla} \times \mathbf{E}^{\prime
}+\frac{1}{c^{\prime }}\frac{\partial \mathbf{B}^{\prime }}{\partial t}=0.
\label{CONVCASE}
\ea
It can be readily verified that the changes  
\ba
&&\mathbf{E}^{\prime }=\sqrt{\epsilon }\mathbf{E}, \qquad \mathbf{B}^\prime=\frac{1}{\sqrt{\mu}} \mathbf{B}, \qquad  c^{\prime }=\frac{c%
}{\sqrt{\mu \epsilon }}\qquad \mathbf{\;}\rho ^{\prime }=\frac{\rho }{\sqrt{%
\epsilon }},\qquad \mathbf{J}^{\prime }=\frac{1}{\sqrt{\epsilon }}\mathbf{J}, \nonumber \\
&& \mathbf{b}^\prime=\sqrt{\frac{\mu}{\epsilon}} \,  \mathbf{b}, \qquad b_0^\prime=\mu \, b_0,
\label{TRANSF}
\ea
produce Maxwell equations for a chiral medium with arbitrary $\epsilon$ and $\mu$
\ba
&&\epsilon \gv{\nabla} \cdot \mathbf{E}=\frac{4\pi}{c}J^{0}-\mathbf{\ b}\cdot \mathbf{B},\qquad \frac{1}{\mu}\gv{\nabla} \times \mathbf{B}- \frac{1%
}{c} \epsilon\frac{\partial \mathbf{E}}{\partial t}=\frac{4\pi}{%
c}\mathbf{J}+{\ b}_{0}\mathbf{B}+%
\mathbf{\ b}\times \mathbf{E}, \nonumber \\
&&
\gv{\nabla} \cdot \mathbf{B}=0,\qquad \gv{\nabla} \times \mathbf{E}+\frac{1}{c}\frac{\partial \mathbf{B}}{\partial t}=0.
\label{CHIRALCASE}
\ea


\end{document}